\documentclass{aa}
\usepackage{times,mathptm}
\usepackage{natbib}
\bibpunct[, ]{(}{)}{;}{a}{}{,}
\usepackage{graphicx}

\begin{document}


\newcommand{\be}{\begin{equation}}
\newcommand{\ee}{\end{equation}}
\newcommand{\bd}{\begin{displaymath}}
\newcommand{\ed}{\end{displaymath}}
\newcommand{\bea}{\begin{eqnarray}}
\newcommand{\eea}{\end{eqnarray}}
\newcommand{\etal}{et al.}

\newcommand{\mum}{\,\mu\hbox{m}}
\newcommand{\mm}{\,\hbox{mm}}
\newcommand{\cm}{\,\hbox{cm}}
\newcommand{\m}{\,\hbox{m}}
\newcommand{\km}{\,\hbox{km}}
\newcommand{\AU}{\,\hbox{AU}}
\newcommand{\second}{\,\hbox{s}}
\newcommand{\yr}{\,\hbox{yr}}
\newcommand{\g}{\,\hbox{g}}
\newcommand{\kg}{\,\hbox{kg}}
\newcommand{\rad}{\,\hbox{rad}}
\newcommand{\erg}{\,\hbox{erg}}


\title{On the nature of clumps in debris disks}

\bigskip
\author{Alexander V. Krivov \inst{1}
        \and
        Martina Queck\inst{1}
        \and
        Torsten L\"ohne\inst{1}
        \and
        Miodrag Srem{\v c}evi{\'c}\inst{2}
       }
\offprints{A.V.~Krivov, \email{krivov@astro.uni-jena.de}}
\institute{Astrophysikalisches Institut, Friedrich-Schiller-Universit\"at Jena,
           Schillerg\"a{\ss}chen~ 2--3, 07745 Jena, Germany
           \and
           LASP, University of Colorado,
           1234 Innovation Drive, Boulder, CO 80303, USA
          }
\date{Received May 10, 2006; accepted September 20, 2006}

\abstract
{
The azimuthal substructure observed in some debris disks, as exemplified by
$\epsilon$~Eridani,
is usually attributed to resonances with embedded planets.
In a standard scenario, the Poynting-Robertson force, possibly
enhanced by the stellar wind drag, is responsible for the  delivery of dust from
outer regions of the disk to locations of external mean-motion planetary
resonances; the captured particles then create characteristic ``clumps''.
Alternatively, it has been
suggested that the observed features in systems like $\epsilon$~Eri
may stem from populations of planetesimals that have been captured in resonances
with the planet, such as Plutinos and Trojans in the solar system.
A large fraction of dust produced by these bodies would stay locked in the same resonance, 
creating the dusty clumps. 
To investigate both scenarios and their applicability limits for a wide range of stars,
planets, disk densities, and planetesimal families we construct simple analytic models for both 
scenarios.
In particular, we show that the first scenario works for disks with the pole-on optical depths
below about $\sim 10^{-4}$--$10^{-5}$.
Above this optical depth level,
the first scenario will generate a narrow resonant ring with a hardly visible
azimuthal structure, rather than clumps.
It is slightly more efficient for more luminous/massive
stars, more massive planets, and planets with smaller orbital radii,
but all these dependencies are weak.
The efficiency of the second scenario is proportional to the mass of the
resonant planetesimal family,
as example, a family with a total mass of $\sim 0.01$ to $0.1$ Earth masses could be 
sufficient to account for the clumps of $\epsilon$~Eridani.
The brightness of the clumps produced by the second scenario increases with 
the decreasing luminosity of the star, increasing planetary mass, and decreasing orbital radius
of the planet. All these dependencies are much stronger than in the first scenario.
Models of the second scenario are quantitatively more uncertain than those of the first one,
because they are very sensitive to poorly known properties 
of the collisional grinding process.

}

\authorrunning{Krivov et al.}
\titlerunning{Clumps in debris disks}

\maketitle

\section{Statement of the problem}

Various kinds of structures~--- offsets, wing asymmetries, warps, voids, clumps, rings,
spirals, inner gaps~--- are either seen in the images of resolved debris disks
or have been retrieved from analyses of spectral energy distributions.
Accordingly, a variety of mechanisms have been suggested to explain the structure formation
\citep[see, e.g.][for a recent review]{augereau-2004}.
Embedded planets can create structures
through secular (long-term) perturbations \citep{mouillet-et-al-1997},
by resonant perturbations \citep{liou-zook-1999},
as well as by gravitational scattering of the disk particles \citep{moromartin-malhotra-2005}.
Some features may stem from recent stellar flybys \citep{kalas-et-al-2001};
others may reflect recent major collisional events between large planetesimals 
\citep{wyatt-dent-2002,kenyon-bromley-2005, grigorieva-et-al-2006}.

Here we address the azimuthal substructure
seen in several disks, best exemplified by
the 0.85~Gyr-old K2V star,
$\epsilon$~Eridani.
Submillimeter images obtained with JCMT/SCUBA
sighted a few bright blobs or clumps in this nearly pole-on disk
\citep{greaves-et-al-1998}.
New observations of the same disk \citep{greaves-et-al-2005}
revealed signs of the disk rotation, demonstrating that three out of six clumps are real
and indicating that the observed rotation rate is compatible with
the Keplerian velocity at the location of the clumps.
A similar clumpy structure has recently been found in the pole-on disk
around another 1~Gyr-old K star, HD~53143 \citep{kalas-et-al-2006}.
Clumps have also been reported for other disks seen at moderate inclinations,
most notably Fomalhaut \citep{holland-et-al-2003}
and Vega \citep{holland-et-al-1998,wilner-et-al-2002}.

It is natural to attribute the formation of blobs to the resonant perturbations
exerted on dust grains by a planet. But how can the resonances create the clumps?
One mechanism, originally proposed by \citet{gold-1975}
and discussed very intensively in the past decade, which we call
{\em scenario~I}, can be summarized as follows.
Dust particles in dilute debris disks are known to steadily migrate inwards
due to the Poynting-Robertson (P-R) force, possibly enhanced by the stellar wind
\citep[e.g.][]{weidenschilling-jackson-1993,beauge-ferrazmello-1994,liou-zook-1997}.
If there is a planet in the disk, the particles reach locations of exterior mean-motion 
resonances (MMRs) with the planet and get trapped, yielding
characteristic density patterns.

\citet{liou-et-al-1996} and \citet{liou-zook-1999} applied this idea to the solar system.
By modeling perturbations induced by jovian planets on the Edgeworth-Kuiper Belt dust disk,
they found  efficient resonant trapping of dust by Neptune
(which produces arcs of dust co-orbital with the planet) and
efficient ejection of dust out of the solar system by Jupiter and Saturn.
Were the solar system observed from outside,
the presence of at least Neptune and Jupiter would be obvious merely from analyzing
images of the dust disk.
While the large-scale clumps on the outskirts of the solar system still escape observational
detection, there is one, so far the only, case where scenario~I is observed at work: the
asymmetric resonant ring of asteroidal dust around the Earth orbit. Predicted by 
\citet{jackson-zook-1989}, it was
identified in IRAS \citep{dermott-et-al-1994} and COBE/DIRBE data \citep{reach-et-al-1995}.
Search of a similar ring around the Mars orbit was not successful, however \citep{kuchner-et-al-2000}.

In the case of $\epsilon$~Eri, 
\citet{liou-et-al-2000},
\citet{ozernoy-et-al-2000},
\citet{quillen-thorndike-2002},
and \citet{deller-maddison-2005}
have tried to find particular orbital parameters and masses of planet, or planets,
that may reproduce the observed substructure in the disk.
A systematic overview of resonant structures that a single planet can form in a debris
disk through this mechanism was painted by \citet{kuchner-holman-2003}.

A major problem with this mechanism is that catastrophic grain-grain collisions
may smear out the planet-induced structure already at moderate optical depths,
because the collisional lifetimes appear to be shorter than the P-R
timescale that determines the dust injection rate into the resonance
\citep{desetangs-et-al-1996b, lagrange-et-al-2000}.
\citet{wyatt-2003,wyatt-2006} suggest an alternative scenario, hereinafter {\em scenario~II},
in which the clumps stem from planetesimals that have been captured
in planetary resonances, possibly as a result of the planet's migration.
Such resonant populations have long been known in the solar system~---
Plutinos are locked in the 3:2 MMR with Neptune, while Greeks and Trojans reside in the primary 
resonance with Jupiter.
A large fraction of the dust which these bodies produce may stay locked in the same resonance
and could well account for the observed clumps.

In this paper, we make a comparative analysis of both scenarios and
estimate typical timescales and efficiency of both scenarios
with simple analytic models. The
goal is to investigate and compare the efficiency of both scenarios for a wide range
of essential parameters, such as the mass of the central star, mass and orbital
radius of the embedded planet, as well as  the optical depth of the ``background'' disk
(in the first scenario) and properties of the resonant family of planetesimals
(in the second one).
Scenarios~I and ~II are considered in Sects.~2 and 3, respectively.
Section 4 contains our conclusions and a discussion.

\section{Scenario~I}

We start with an analysis of scenario~I, in which dust particles
are brought from afar to resonance locations by the P-R effect and are captured
in resonances.
A resonance implies a $(p+q):p$-commensurability of the mean motions
of the planet and the grain.
If $a_p$ is the semimajor axis of the planet orbit, the resonant value of the 
grain's orbital semimajor axis is
\be \label{a res}
  a = a_{p}(1-\beta)^{1/3}\left( p+q \over p \right)^{2/3},
\ee
with $\beta$ being the radiation pressure to gravity ratio for the dust grain
(assumed to be a perfect absorber):
\be
  \beta
  =
  0.57
  \left( L_\ast \over L_\odot \right)
  \left( M_\odot \over M_\ast \right)
  \left( 1\g\cm^{-3} \over \rho \right)
  \left( 1\mum \over s \right) ,
\ee
where $L_\ast/L_\odot$ and $M_\ast/M_\odot$ are luminosity and mass of the star in solar
units and $s$ is the grain radius.
In what follows, we treat $M_\ast$ as a parameter and calculate luminosity by using
the standard $L_\ast \propto M_\ast^4$ relation for main-sequence stars.
Throughout the paper, the bulk density of $2\g\cm^{-3}$ is adopted.
To keep our treatment simple, we confine our analysis to
a circular planetary orbit.
The typical resonant dynamics can then be summarized as follows
\citep[e.g.][]{weidenschilling-jackson-1993,beauge-ferrazmello-1994,liou-zook-1997}.
When trapped, a dust grain preserves the semimajor axis.
If $\lambda$ and $\lambda_P$ are longitudes of the grain and planet, respectively,
and $\tilde\omega$ is the longitude of pericenter of the grain orbit,
the resonant argument
\bea
  \Phi = (p+q)\lambda -p\lambda_P - q\tilde{\omega} 
  \label{Phi(lam,om)}
\eea
librates with an amplitude $A$ around the value $\approx 180^\circ$
\citep[cf.][]{kuchner-holman-2003,wyatt-2003,wyatt-2006}.
The eccentricity gradually pumps up. For initially circular orbits
\citep{liou-zook-1997}
\be \label{e(t)}
  e^2(t) = {q\over 3p}\left[ 1 - \exp\left( - {3 p \over p+q} \, B \, t \right)\right],
\ee
where
\be
  B = {2 GM_\ast \beta \over c a^2},
\ee
$GM_\ast$ standing for the gravitational parameter of the central star and
$c$ for the speed of light.
When the eccentricity reaches a certain value $e_{res}$,
the particle becomes a planet-crosser
and shortly thereafter either collides with the planet or gets ejected out of the resonance
zone after a close encounter.
The orbital inclination $i$ exponentially decreases on the same timescale as
$e$ increases; here we assume $0 \le i \le \epsilon$, where
$\epsilon$ is a (small) semi-opening angle of the disk.

Note that some grains that used to be trapped in
a resonance may then continue migrating toward the star,
instead of being ejected from the system. This is, however, more typical
of resonances with very low-mass planets~--- e.g. in the Earth resonant ring 
mentioned above. For such planets,
overlapping of first-order resonances and the onset of chaos,
as well as external perturbations by other planets, may also play a role,
making the dynamics more complicated \citep{marzari-vanzani-1994}. 
Since dust structures induced by terrestrial
planets around {\em other} stars are far beyond observational limits,
our analysis is confined to at least Neptune/Uranus-mass planets
and thus ignores any temporary resonant capture of particles.

To construct a simple model, we choose a ``typical'' grain radius $s$ and a corresponding
$\beta$-value.
This is the size of the grains that compromises the cross section-dominating size 
and the size for which the probability of capture in the resonance ($p_{res}$ below)
becomes significant.
The former corresponds to $\beta$ of several tenths 
\citep{krivov-et-al-2000b,thebault-et-al-2003,krivov-et-al-2006}.
The latter requires $\beta \la 0.1$ \citep[e.g.][]{liou-et-al-1996,liou-zook-1999};
the smaller $\beta$, the larger $p_{res}$.
In our model, we  treat $\beta$ as a free parameter and use $\beta = 0.1$ as a default value
in numerical examples.

For a certain MMR and a given $\beta$ ratio, the resonant value of 
semimajor axis $a$ is given by Eq.~(\ref{a res}).
The particles drift towards the star by P-R drag.
Assume that $\dot{n}^+$ grains cross $r=a$ per unit time.
Let $p_{res}$ be the probability of trapping in the resonance.
Its value depends on size, initial eccentricity, and the inclination of the grain orbit, 
planet's mass $M_p$, and 
varies from one resonance to another 
\citep[e.g.][]{weidenschilling-jackson-1993,beauge-ferrazmello-1994,liou-et-al-1996,liou-zook-1997}. 
For instance, \citet{liou-zook-1999} report that $p_{res}$  increases from
about 0.2 for $\beta=0.4$ to
about 0.5 for $\beta=0.05$
for the Edgeworth-Kuiper belt particles' trapping by Neptune.
Inside a certain range of these parameters, however, $p_{res}$ is not {\em very}
sensitive to their variation. We performed a number of numerical integrations, which we do not
describe here, because the technique is very well known
\citep[e.g.][]{weidenschilling-jackson-1993,beauge-ferrazmello-1994,liou-zook-1997,%
quillen-thorndike-2002,deller-maddison-2005}.
For the parameters of the $\epsilon$~Eri system, with initial eccentricities $\la 0.1$, 
inclinations below $10^\circ$, and
a planet at $40\AU$ with 0.1 Jupiter mass,
the trapping probability is between 10\% and almost 100\% for many resonances
(2:1, 3:2, 8:5, 5:3, 9:5, and others).
In our model, $p_{res}$ is a parameter. In numerical examples,
$p_{res} = 0.5$ is used as a default value.

Besides $p_{res}$, another quantity we need is $e_{res}$,
the value of orbital eccentricity that the captured grains have to 
develop to quit the resonance.
On the basis of the papers cited above and our own numerical integrations,
the typical values lie between 0.1 and 0.4.
Like $\beta$ and $p_{res}$, $e_{res}$ is a model parameter.
As a ``standard'' value, we adopt $e_{res}= 0.2$.
Note that grains do not necessarily reach $e = e_{res}$.
Let us denote by $T_{res}$ 
the time interval in which $e_{res}$ is attained.
As the particles are lost not only because they pump up the orbital eccentricities
to the resonance-quitting value $e_{res}$, but also are eliminated by mutual collisions,
their mean actual lifetime $T$ is shorter than $T_{res}$.
Therefore, their actual maximum eccentricity $e_{max}$,
i.e. the maximum eccentricity achieved
by the particles in their lifetime $T$,
will be smaller than $e_{res}$.

\begin{figure}
  \begin{center}
  \includegraphics[scale=0.90]{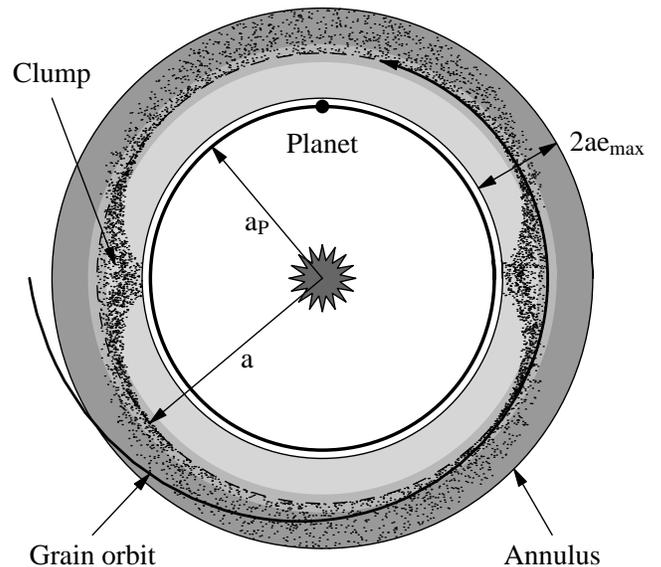} 
  \end{center}
  \caption{
  Schematic of the resonant and non-resonant components of the dust disk
  in the vicinity of a given resonance.
  The particular resonant population with two clumps shown here corresponds to the 
  3:2-MMR with $e_{max}=0.2$ and $A = 20^\circ$.
  }
  \label{fig_cartoon}
\end{figure}

We now introduce an annulus in the disk around the location of a given MMR
(Fig.~\ref{fig_cartoon}), defined as follows.
We require that the annulus width matches the maximum radial excursion of the trapped particles:
from $a(1+e_{max})$ to $a(1-e_{max})$.
At any instant in time, the annulus contains two populations of particles.
One is a
non-resonant ``background'' population of particles,
including those that have not reached the resonance location yet,
as well as those that drifted through it without getting trapped.
It might also include grains that continue drifting
toward the star after quitting resonance. If such grains are present
in sufficient amounts, $p_{res}$ should be re-defined to account for them.
All these particles have a uniform azimuthal distribution.
Another population is
an unevenly distributed (clumpy) population of resonant grains.
Hereinafter, all quantities with subscript ``0'' characterize the background
population and those without subscript the resonant population.
For instance, $n_0$ and $n$ stand for the total number of background grains
and resonant grains in the annulus, respectively.

We now construct a simple kinetic model.
By applying any standard method of the kinetic theory, or simply
by ``counting'' the particles injected into and destroyed in the annulus
per unit time, one finds the balance equation:
\be \label{balance}
  {dn \over dt} = \dot{n}^+ p_{res} - {n \over T_{res}} 
  - {n \over T_{coll}^0} 
  - {n \over T_{coll}} ,
\ee
where $T$s are timescales as explained below.
As $n$ does not have a grain size/mass as argument,
Eq.~(\ref{balance}) can be referred to as the Boltzmann equation.
Since $n$ does not have any grain velocity argument either,
it can also be termed the Smoluchowski equation.
See, e.g., \citet{krivov-et-al-2005} and references therein
for a detailed discussion of a general form of the balance equation.

The resonant population implies several natural timescales associated with 
the terms on the right-hand side of Eq.~(\ref{balance}).
The first is $T_{PR}$, the P-R time to drift from  $a(1+e_{max})$ to $a(1-e_{max})$.
The second is $T_{res}$, the time of eccentricity pumping to the value $e_{res}$ for 
trapped grains.
It can also be treated as the maximum time of residence in the resonance for trapped grains.
Finally, $T_{coll}^0$ and $T_{coll}$
are the lifetimes of the resonant grains against collisions with
background grains and other resonant grains, respectively.

The actual lifetime of grains as determined by all sinks together is given by
\be \label{lifetime}
  {1 \over T} 
  \equiv
    {1 \over T_{res}} 
  + {1 \over T_{coll}^0} 
  + {1 \over T_{coll}} 
\ee
and the maximum 
eccentricity achieved
by the particles in the time $T$, as follows
from Eq.~(\ref{e(t)}) for small $e_{max}$, is
\be \label{e_max}
  e_{max} \equiv e(T)
  \approx
  \sqrt{ {q \over p+q} B T }.
\ee

We return to Eq.~(\ref{balance}) and begin with its first term.
The steady-state number of non-resonant grains in the annulus, $n_0$, is
\be \label{n0}
 n_0 = \dot{n}^+ \left(1 - {p_{res}\over 2} \right) T_{PR},
\ee
where the P-R drift time is given by \citep[e.g.][]{burns-et-al-1979}
\be \label{T_PR}
  T_{PR}
  = {4 e_{max} \over B} .
\ee
Equation~(\ref{n0}) reflects the fact that the outer half of the annulus,
the region between  $a(1+e_{max})$ and $a$, is always filled by the
background grains, regardless of whether some of them get captured in
the resonance upon arrival at $a$. In contrast, the number of grains
in the inner half, between $a$ and $a(1-e_{max})$, lies between
zero (if $p_{res}=1$) and the number of grains in the outer half (if $p_{res}=0$).
These two regions are shown in Fig.~\ref{fig_cartoon} with dark and light grey,
respectively. We note, however, that this division of the annulus into
the outer and inner parts is done in terms of the semimajor axis and not in
real space.
Since grain orbits have non-zero eccentricities, no sharp boundary between
the two halves exists in terms of distance, i.e. at $r=a$.
The boundaries of the whole annulus will also be somewhat fuzzy.

We introduce the normal optical depth $\tau_0$ of the non-resonant population
in the {\em outer} part of the annulus, which is not affected by the resonance.
It is a product of the number of grains in  the outer part and the cross section
of a grain, divided by the area of the outer half of the annulus:
\be \label{tau0}
\tau_0 = {\dot{n}^+ T_{PR}/2 \cdot \pi s^2
         \over
         2 \pi a^2 e_{max} } .
\ee
Substituting Eq. (\ref{T_PR}) and solving for $\dot{n}^+$ results in
\be \label{n+}
  \dot{n}^+
  = B \; \left(a \over s \right)^2\; \tau_0 .
\ee
Thus $\dot{n}^+$ and thereby the first term in Eq.~(\ref{balance}) are
expressed through an ``observable'' quantity, the background optical depth $\tau_0$.

The resonant pumping time $T_{res}$ in the second term of Eq.~(\ref{balance})
comes from Eq.~(\ref{e(t)}). For small $e_{res}$,
\be \label{T_res}
  T_{res} \approx {p+q \over q} {e_{res}^2 \over B} .
\ee

Finally, the collisional lifetimes $T_{coll}^0$ and  $T_{coll}$ 
in Eq.~(\ref{balance}) should be estimated.
General expressions are
\be \label{T_coll_both}
  T_{coll}^0 = { V_0 \over n_0 \; \sigma \; v_0 }
\qquad
{\rm and}
\qquad
  T_{coll} = { V \over n \; \sigma \; v } ,
\ee
where $\sigma = 4 \pi s^2$ is the collision cross section of the like-sized particles,
$v_0$ the mean impact velocity between background and resonant grains,
$V_0$  the effective volume of their interaction specified below,
and $v$ and $V$ are similar quantities for mutual collisions of resonant grains.

Consider $T_{coll}^0$ and the third term of Eq.~(\ref{balance}) first.
The interaction volume
is given by \citep[see, e.g., Appendix in][]{krivov-et-al-2006}
\be
  V_0 = (2 \pi a)^2 h |\sin\gamma| ,
\ee
where $h$ is the disk thickness ($h = 2 a \sin \epsilon$ for a disk with the semi-opening
angle $\epsilon$) and $\gamma$  the angle between
the orbits of the particles locked in the resonance
and the orbits of non-resonant
grains at their intersection point.
With
\be \label{v kepler}
  v_k = (GM_\ast(1-\beta)/a)^{1/2}
\ee
being the Keplerian circular velocity
and neglecting corrections due to small eccentricities and inclinations,
we have
\be \label{v_0}
  v_0 \approx v_k |\sin\gamma|.
\ee
Thus the ratio $V_0/v_0$ in the first Eq.~(\ref{T_coll_both}) is independent of
eccentricities and inclinations, as long as they are small, and
that equation simplifies to
\be \label{T_coll^0}
  T_{coll}^0 = {1 \over n_0} {(2 \pi a)^2 h \over \sigma v_k}
           \equiv {1 \over n_0} \hat{T}_{coll} .
\ee
We also note that, in the monosize model considered, $T_{coll}^0$ is nearly independent
of the particle size chosen.
This is seen from
Eqs.~(\ref{n0}) and (\ref{tau0}),
if we neglect  a weak dependence of $v_k$ on the grain radius.

To compute $T_{coll}$  for the fourth term of Eq.~(\ref{balance}), we
need the interaction volume $V$ and the average impact velocity $v$ for
mutual collisions between the particles in the resonant clumps.
Accurate calculations for this case are not easy,
because the resonant population has a distribution of eccentricities from $0$ to $e_{max}$
and a highly non-uniform distribution of apsidal lines.
There is no guarantee therefore that the particle-in-a-box-like estimates
we employed for $T_{coll}^0$  would
yield correct results.
A full solution of this problem, which may have a number of applications to resonant systems,
is beyond the scope of this paper.
Our calculations (Queck et al., in prep.)
show, however, that an analog of Eq.~(\ref{T_coll^0}),
\be \label{T_coll}
  T_{coll} = {1 \over n} {(2 \pi a)^2 h \over \sigma v_k}
           = {1 \over n} \hat{T}_{coll} ,
\ee
provides results that are within a factor of two from the ``true'' values.

Thus all characteristic timescales, and all four terms on the right-hand 
side of  Eq.~(\ref{balance}), are now determined.
A little technical complication arises from the fact that
$T_{coll}^0$ in Eq. (\ref{T_coll^0}) depends through $n_0$ on $e_{max}$
(Eqs.~\ref{n0}--\ref{T_PR}) which, in turn, depends on $T$ and thus back on $T_{coll}^0$
(Eqs.~\ref{lifetime}--\ref{e_max}).
Actual calculations have therefore been done iteratively,
with $e_{res}$ as the first approximation for $e_{max}$.

\begin{figure}
  \begin{center}
  \includegraphics[scale=0.32]{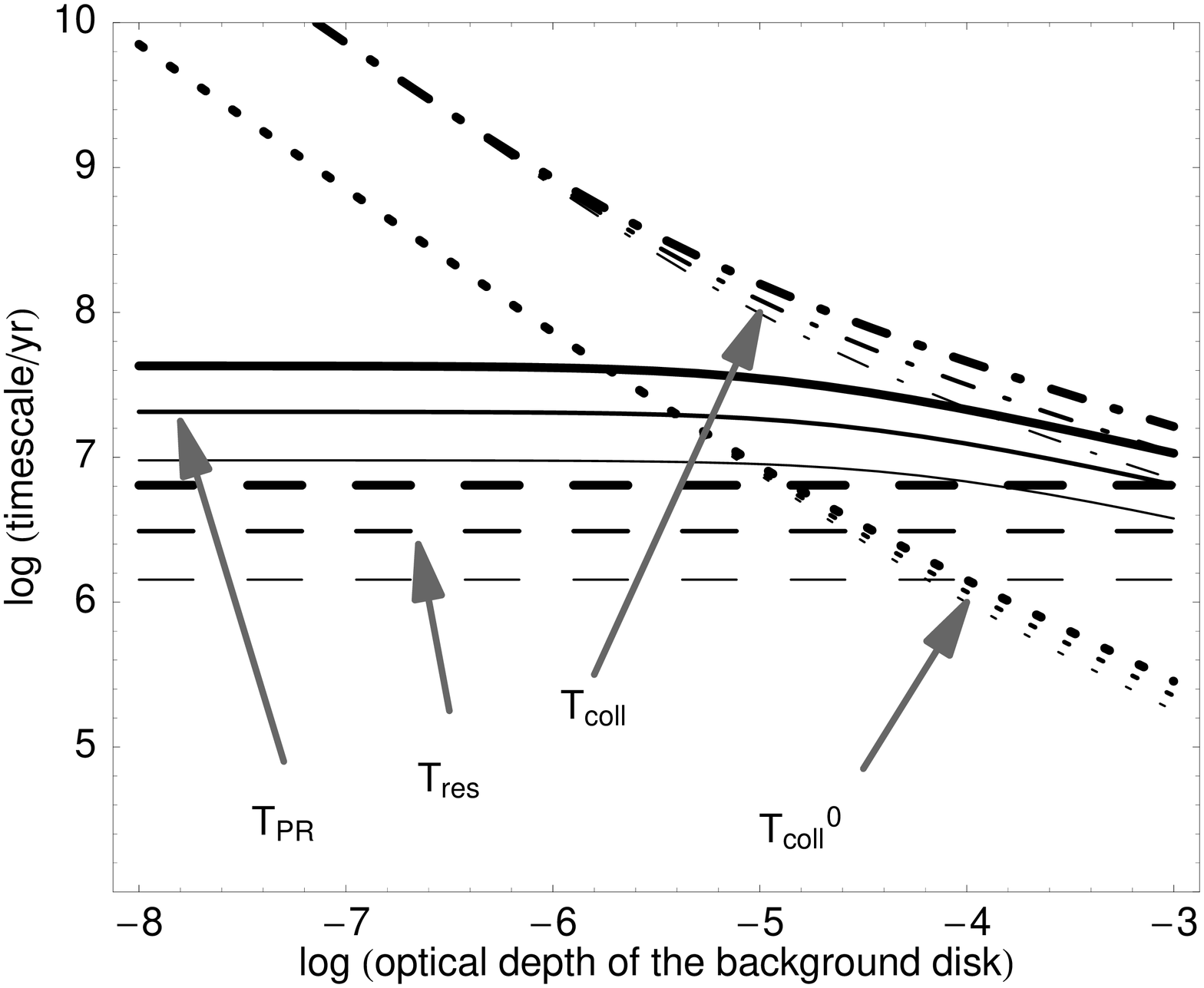} 
  \end{center}
  \caption{Typical timescales as functions of the normal optical depth
  $\tau_0$  of the non-resonant disk:
  $T_{PR}$ (solid)
  $T_{res}$ (dashed),
  $T_{coll}^0$ (dotted),
  $T_{coll}$ (dash-dotted).
  Thick, medium, and thin lines: $\beta=0.05$, 0.10, and 0.20, respectively.
  These values correspond to $s=2.9$, $1.5$, and $0.7\mum$.
  The collisional time $T_{coll}^0$ is nearly the same for all three values (see text).  
  }
  \label{fig_timescales}
\end{figure}

For illustrative purposes, we choose the parameters suitable for the
$\epsilon$ Eridani disk:
$M_\ast=0.8M_\odot$,
the presumed planet at $a_p=40\AU$
\citep{liou-et-al-2000,ozernoy-et-al-2000,quillen-thorndike-2002,deller-maddison-2005}.
The collisional lifetime (\ref{T_coll^0}) is proportional to the semi-opening angle
of the disk $\epsilon$, which is unconstrained. By analogy with other
debris disks seen edge-on ($\beta$~Pic, AU~Mic), we take
$\epsilon=0.1=6^\circ$.
We choose 3:2-MMR, but the results are very similar for other low-degree, low-order
resonances.
Figure~\ref{fig_timescales} plots the timescales as functions of the optical
depth $\tau_0$ of the background disk,
showing that the collisional lifetime $T_{coll}^0$
is shorter than the other timescales
for disks with roughly $\tau_0 \ga 10^{-5}$, which is the case for $\epsilon$~Eri.

\begin{figure}
  \begin{center}
  \includegraphics[scale=0.32]{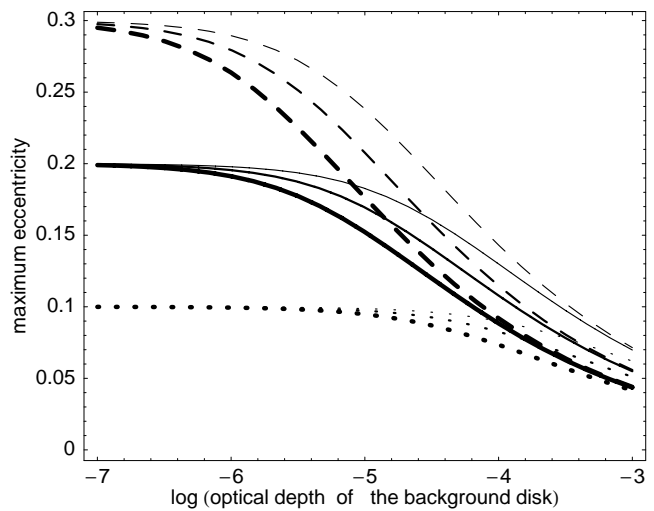} 
  \end{center}
  \caption{
  Maximum orbital eccentricity of the clump particles
  as a function of the optical depth of the ``regular'' disk $\tau_0$.
  Dashed, solid, and dotted lines:
  $e_{res}=0.3$, $0.2$, and $0.1$, respectively.
  Thick, medium, and thin lines correspond to
  $\beta=0.05$, 0.10, and 0.20.
  }
  \label{fig_scenarioI_emax}
\end{figure}

Figure~\ref{fig_scenarioI_emax} shows how the maximum orbital eccentricity
$e_{max}$ of resonant particles decreases with increasing optical depth.
It also shows that, for denser disks, $e_{max}$ depends only weakly on the
maximum value $e_{res}$ determined by the resonant dynamics.
For $\tau_0 \sim 10^{-4}$, the resulting value $e_{max}$ is close to 0.1,
regardless of $e_{res}$.
For such low eccentricities, the resonant population will look like a narrow 
ring without azimuthal structure, not like clumps. The dustier the disk, the
narrower the resulting ring (for planets in near-circular orbits),
and the less pronounced the azimuthal structure.
Therefore, prior to any calculation of the clumps' optical depth,
we can conclude that in sufficiently dusty disks,
scenario~I fails to produce the clumps.

We now solve the balance equation (\ref{balance}).
The collisional lifetime of resonant particles against mutual collisions,
Eq.~(\ref{T_coll}), depends on $n$, which makes the whole equation
Eq.~(\ref{balance}) quadratic in~$n$.
At $t \rightarrow \infty$, the total number of resonant particles tends to
\be  \label{quadratic}
n = {\hat{T}_{coll} \over 2}
    \left[
      \sqrt{
               { 4 \dot{n}^+ p_{res} \over \hat{T}_{coll} }
               +
               \left(
                 {1 \over T_{res}} + {1 \over T_{coll}^0}
               \right)^2
             }
       -
       \left(
         {1 \over T_{res}} + {1 \over T_{coll}^0}
       \right)
    \right] .
\ee
In the limit $n \ll n_0$, meaning relatively faint clumps
and a relatively bright ``regular'' disk,
the fourth term of Eq.~(\ref{balance}) is much smaller than
the third one, so that Eq.~(\ref{balance}) is linear in $n$.
In this case, the above solution simplifies to
\be  \label{linear}
  n  = { \dot{n}^+ p_{res}
         \over
         T_{res}^{-1} + (T_{coll}^0)^{-1}
       }.
\ee

It remains to estimate the optical depth of the resonant clumps.
The clumps are not expected to be uniform for several reasons.
First, even if the orbital eccentricities of all grains were equal,
the second Kepler's law and the synodic motion of the particles
relative to the planet would make the surface density of the resonant population
non-uniform.
Second, the resonant population consists of grains with a distribution
of eccentricities, which is not uniform either:
orbital dynamics forces the particle to
spend most of the time in orbits with higher eccentricities
up to $e_{res}$ (see. Eq.~\ref{e(t)}), whereas mutual collisions
tend to  eliminate these long-lived grains, shifting the distribution
towards $e \ll e_{res}$. A superposition of these effects makes a spatial
distribution of the optical depth in the clumps rather complex
and dissimilar for different resonances
and libration amplitudes $A$.

To compute the normal optical depth $\tau$ of the clumps, we use the following
approximate method.
We multiply $n$ by the cross-section area of a particle, $\pi s^2$, and divide
the result by an approximate area of the clumps $S$. Noting that
the area of the annulus is $S_0 = 4 \pi a^2 e_{max}$,
we introduce the fractional area occupied by the clumps,
$\hat{S} \equiv S/S_0$.
Then,
\be
\tau \approx {n s^2 \over 4 \hat{S} a^2 e_{max} } .
\ee
The fractional area $\hat{S}$
was determined with the aid of a straightforward numerical simulation,
by directly counting the points on the scatter plots.
We considered the 3:2 resonance again, assuming different values of 
$e_{max}$ between zero and $e_{res}=0.2$, as well as different values of $A$.
For $e_{max}=0.2$,
taking into account {\em all} spatial locations reached by the particles
locked in the resonance leads to $\hat{S}$ 
from $\approx 0.5$ for $A = 20^\circ$
to $\approx 0.8$ for $A = 90^\circ$.
Defining, however, the regions with 30\% of the peak brightness as a clump boundary,
we arrive at values $\approx 0.1$ for $A = 20^\circ$
to $\approx 0.3$ for $A = 90^\circ$.
Dependence on $e_{max}$ turned out to be rather weak:
$\hat{S}$ grows slightly with decreasing $e_{max}$.
We therefore adopt a simple $\hat{S} = 0.2$
in our numerical examples, and stress that
the actual $\tau$ in the brightest portions of the clumps is
several times larger than we predict.

\begin{figure}
  \begin{center}
  \includegraphics[scale=0.32]{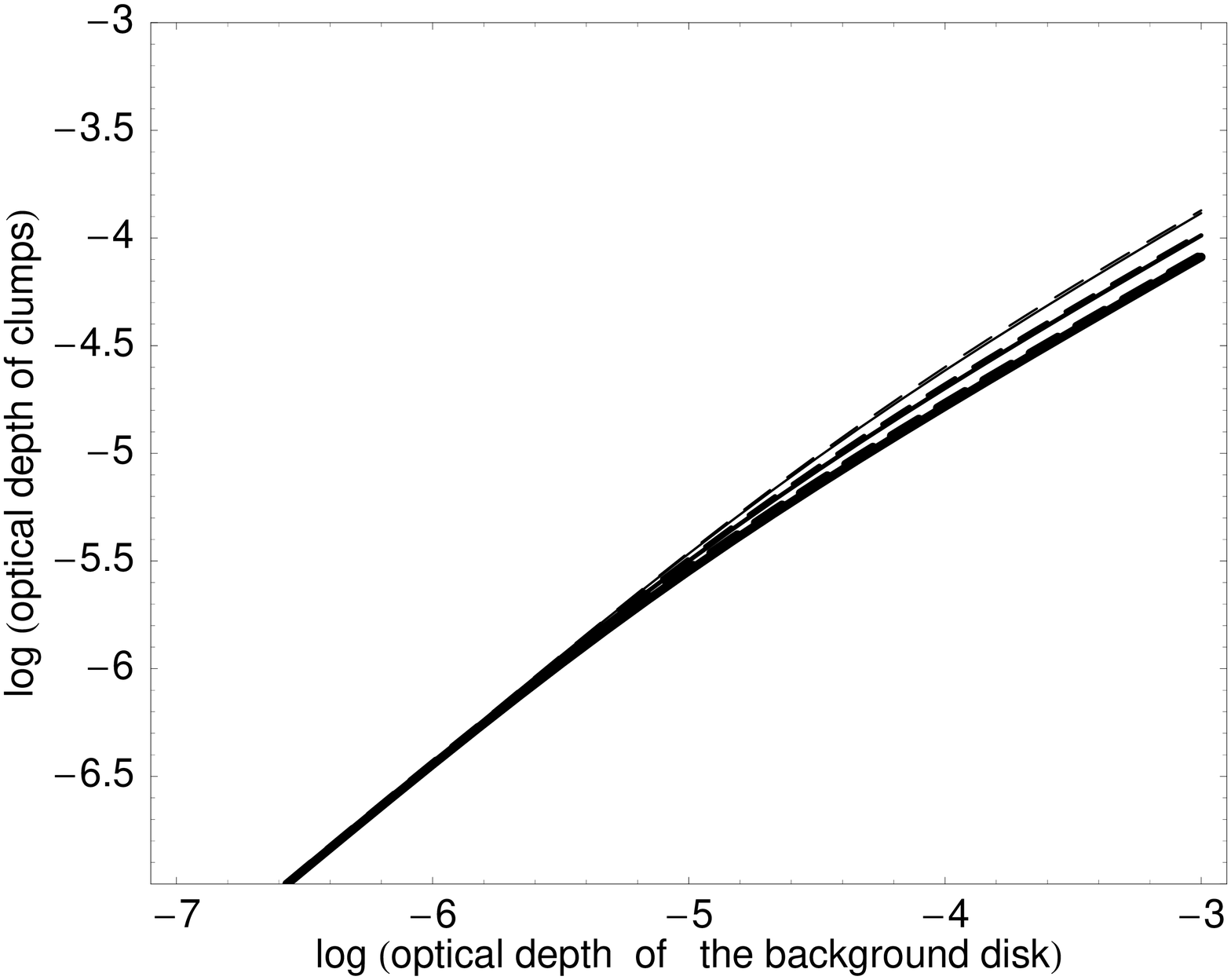}\\
  \includegraphics[scale=0.32]{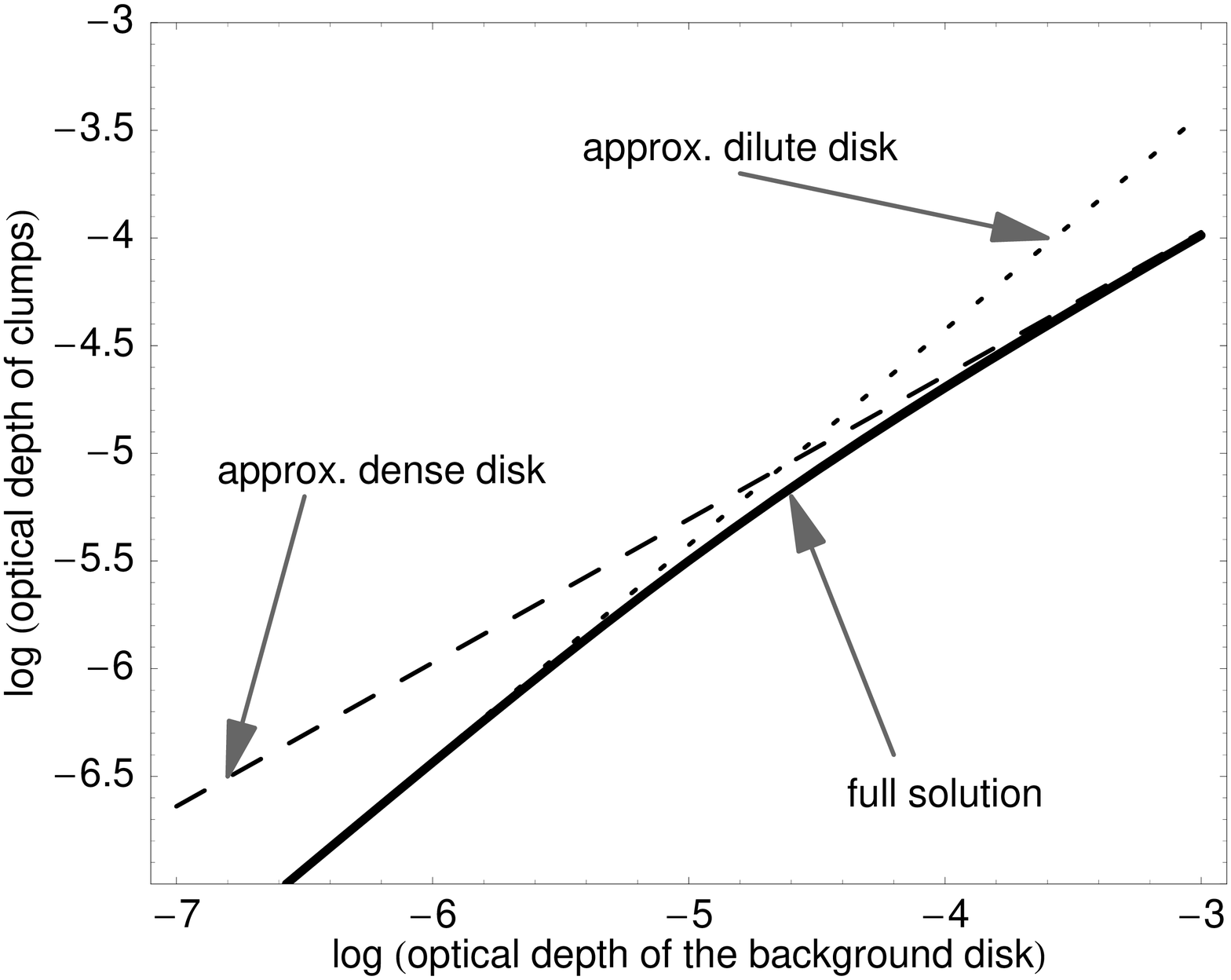}\\
  \end{center}
  \caption{
  Normal optical depth of the resonant clumps $\tau$ as a function
  of the optical depth of the ``regular'' disk $\tau_0$.
  Top: full solution (\ref{quadratic}) of the balance equation
  (solid lines) and solution (\ref{linear}) of the linearized balance equation
  (dashed lines).
  Thick, medium, and thin lines correspond to
  $\beta=0.05$, 0.10, and 0.20, respectively.
  Bottom: 
  full solution (\ref{quadratic}) of the balance equation for $\beta=0.1$
  (solid line)
  versus the tenuous disk approximation (\ref{tau tenuous}) (dotted)
  and denser disk approximation (\ref{tau dense})(dashed).
  }
  \label{fig_scenarioI}
\end{figure}

Figure~\ref{fig_scenarioI} (top) depicts the resulting optical depth of the resonant blobs
versus optical depth of the ``regular'' disk.
The optical depth of clumps
grows more slowly than that of the non-resonant population.
A comparison between solid lines (Eq.~\ref{quadratic})
and dashed ones (Eq.~\ref{linear}) of the same thickness shows
that the linear approximation (\ref{linear}) provides  very good accuracy.

We now seek an approximate analytic scaling formula
for the optical depth of the clumps,
considering the stellar mass $M_\ast$, the planet's orbital radius $a_p$,
and a number of other quantities as free parameters.
The ratio of the normal optical depth of the clumps, $\tau$, and
of the background disk, $\tau_0$, is
\be \label{tau to tau0 init}
  \frac{\tau}{\tau_0} = {n \over n_0 } \cdot { 1-p_{res}/2 \over \hat{S} } .
\ee
From Eq.~(\ref{n0}) and an obvious relation
\be \label{n}
   n = \dot{n}^+ p_{res} T ,
\ee
we obtain
\be \label{tau to tau0 intermed}
  \frac{\tau}{\tau_0}
  =
  p_{res}
  \cdot
  { T \over \hat{S} T_{PR} },
\ee
where $T_{PR}$ is given by Eqs.~(\ref{T_PR}) and (\ref{e_max}).

For {\em tenuous disks}
(such that $T_{coll} \gg T_{res}$),
$T \approx T_{res}$ and
$e_{max}(T) \approx e_{res}$.
Therefore,
\be
  \frac{\tau}{\tau_0}
  \approx
  p_{res}
  \cdot
  {T_{res} \over \hat{S} T_{PR}}
  = 
  p_{res}
  \cdot
  {B T_{res} \over 4 \hat{S} e_{res}}
\ee
or, with the aid of Eq.~(\ref{T_res}),
\be \label{tau tenuous}
  {\tau \over \tau_0}
  =
  p_{res}
  \cdot  
  {p + q \over q}
  \cdot  
  {e_{res} \over 4 \hat{S} } .
\ee
The same formula in a form that is convenient for numerical estimates is
\be \label{tau tenuous num}
  {\tau \over \tau_0}
  =
  0.125
  \left( p_{res} \over 0.5 \right)
  \left( \hat{S} \over 0.2 \right)^{-1}
  \left( e_{res} \over 0.2 \right)
  \left( p + q \over q \right) .
\ee
This equation shows that $\tau/\tau_0$ is typically smaller
(but not much smaller) than unity.
In some cases it can exceed unity, however.

For {\em denser disks}
(such that $T_{coll} \ll T_{res}$),
the lifetime $T$ in Eq.~(\ref{tau to tau0 intermed})
has to be evaluated iteratively, which 
complicates the calculations considerably.
To keep the treatment at a reasonably simple level,
we therefore assume that $n \ll n_0$,
which is justified by Eq.~(\ref{tau to tau0 init})
and by the fact that
$\tau$ does not exceed $\tau_0$ for sufficiently dusty
disks.
In this case
$T \approx T_{coll}^0 = \hat{T}_{coll}/ n_0$.
We then substitute into
Eq.~(\ref{tau to tau0 intermed}) expressions
(\ref{T_coll}) for $\hat{T}_{coll}$,
(\ref{n0}) for $n_0$,
(\ref{T_PR}) for $T_{PR}$,
and so on  ``to the whole depth'' to express them
through the free parameters.
Omitting straightforward, but lengthy algebra,
the result is
\bea \label{tau dense}
  \tau
  &\approx&
  p_{res}
  \cdot
  \frac{1}{4\hat{S}}
 \nonumber\\
   &\times&
   \left[
     {\beta \pi \sin\epsilon
      \over
      c \left( 1-{p_{res} \over 2} \right)(1-\beta)^{2/3}
     }
     \sqrt{\frac{GM_\ast}{a_p}}
     \tau_0^2
   \right.
 \nonumber\\
     &&\times
   \left.
     \left(\frac{p + q}{q}\right)^2
     \left(\frac{p}{p+q}\right)^{1/3}
   \right]^{1/3},
\eea
where the square-root expression is the planet's orbital velocity.
Dropping factors in brackets that are close to unity,
we obtain a somewhat cruder version
of the same formula which is convenient for numerical estimates:
\bea \label{tau dense num}
  \tau
  &\approx&
  9 \times10^{-6}
  \left( p_{res} \over 0.5 \right)
  \left( \hat{S} \over 0.2 \right)^{-1}
  \left( \beta \over 0.1 \right)
  \left( \epsilon \over 0.1 {\rm rad} \right)
 \nonumber\\
  &\times&
  \left( M_\ast \over M_\odot \right)^{1/6}
  \left( a_p \over 100\AU \right)^{-1/6}
  \left( \tau_0 \over 10^{-4} \right)^{2/3}
  \left( p + q \over q \right)^{2/3} .
\eea
The bottom panel in Fig.~\ref{fig_scenarioI} compares
both analytic approximations, Eqs.~(\ref{tau tenuous}) and (\ref{tau dense}),
with each other and with the ``exact'' solution
(\ref{quadratic}).

\begin{figure*}[htb!]
  \begin{center}
  \includegraphics[scale=0.33]{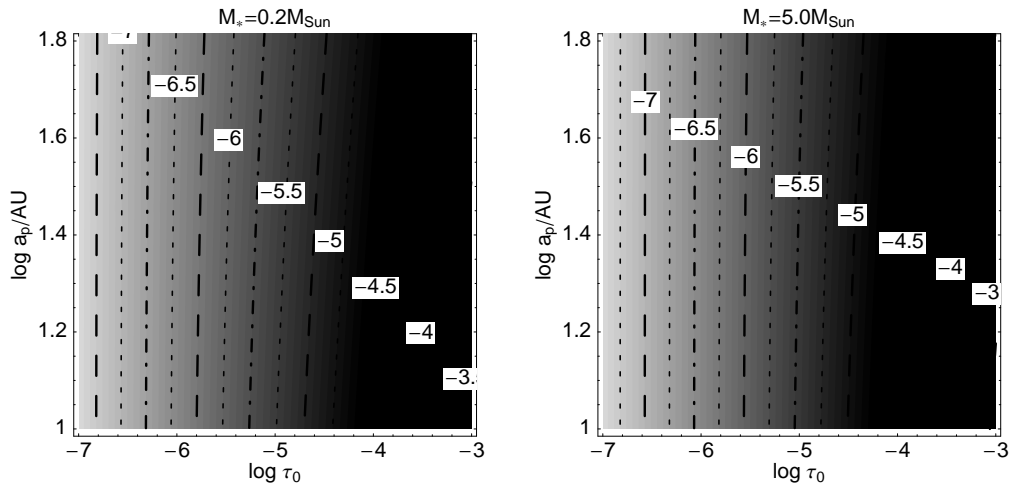} 
  \end{center}
  \caption{
  Contour plots of the normal optical depth of the resonant clumps $\tau$, as a function
  of the optical depth of the ``regular'' disk $\tau_0$ and the planet's distance from the star.
  Left and right: stars with 0.2 and $5.0M_\odot$.
  Contours are labeled with the values of $\log\tau$.
  Assumed $\beta=0.10$.
  }
  \label{fig_scenarioI_plots}
\end{figure*}

Equations~(\ref{tau tenuous}) and (\ref{tau dense})
explicitly show the dependence of $\tau$ on many parameters:
the stellar mass $M_\ast$, the planet's orbital radius $a_p$,
the resonance integers $p$ and $q$,
the typical $\beta$ value of constituent particles,
and others.
For tenuous disks, $\tau$ is nearly proportional to $\tau_0$.
For denser disks, $\tau \propto \tau_0^{2/3}$~--- a ``slowdown''
effect seen before in Fig.~\ref{fig_scenarioI}.
The clumps' optical depth is nearly independent of
the planet's distance from the star $a_p$ for tenuous disks
and slightly decreases with increasing $a_p$ for larger $\tau_0$.
The dependence on the stellar mass $M_\star$ is absent for
dilute disks and weak for denser ones: $\tau$ slightly
increases for more massive primaries.
Note that the last statement is only true if we fix $\beta$
and not $s$. If the grain radius $s$ were treated as a free parameter,
we would have to express $\beta$ through $s$ in Eq.~(\ref{tau dense}),
which would yield an additional $M_\star^{1/3}$ in Eq.~(\ref{tau dense}),
making the dependence on the stellar mass stronger.
Next, an important dependence on
the planet mass $M_p$ 
is implicit (through the capture probability $p_{res}$);
we will discuss it later.
The fractional area of the clumps, $\hat{S}$, depends on the
maximum orbital eccentricity of the resonant particles, and
thus on other parameters such as $\tau_0$, but only weakly.
For denser disks, $\tau$ increases, albeit only slightly,
with the semi-opening angle of the background disk.

Figure \ref{fig_scenarioI_plots} presents
the ``exact'' solution, based on Eq.~(\ref{quadratic}),
in the form of contour plots
of equal $\tau$, the average optical depth of the clumps.
The figure shows that the phenomenon seen in Fig.~\ref{fig_scenarioI}~---
``slowdown'' of the growth of $\tau$
at higher $\tau_0$~--- holds for all distances from the star $a_p$.
The effect is a bit more pronounced at larger $a_p$.
A comparison of the left and right panels
demonstrates another effect: for given $\tau_0$ and $a_p$, disks around
more massive/luminous stars can develop somewhat denser clumps.
All these effects are almost negligible, however.

As for any analytic model, our model rests on a number of simplifying assumptions,
which we now discuss.
One of those is a constant value of the resonance trapping probability $p_{res}$,
which we have set to 0.5 in the numerical examples.
In fact, it is through this quantity that the 
planet mass
$M_p$ would enter the model. This probability also
depends on a particle's $\beta$, orbital semimajor axis $a$,
and it varies from one resonance to another. Non-zero inclinations can affect the
trapping probability as well.
To the best of our knowledge, analytic or empirical
formulas for $p_{res}$ as a function of $M_p$ are only available for some subspaces
of these parameters, and only for the planar problem
\citep{beauge-ferrazmello-1994,lazzaro-et-al-1994}.
It is clear, however, that resonance trapping is only possible within a certain
planet mass range. If the planet is too massive, the orbits of dust grains become
chaotic \citep{kuchner-holman-2003}.
For a solar-mass star, the maximum mass is typically several Jupiter masses.
If, conversely, the planet mass is too low, the particles go through the resonance
without getting trapped.
\citet{beauge-ferrazmello-1994} investigated the problem analytically and found,
for instance, $0.04M_J$ ($M_J =$ Jupiter mass) to be
a minimum mass to trap grains with $\beta=0.1$ in a 3:2 MMR.
In a more general study, \citet{lazzaro-et-al-1994} found that for a planet with
$M_p \ga 0.01M_J$ and for micrometer-sized particles with eccentricities lower than
a few percent, there are large regions of capture in phase diagrams.
These analytic estimates are supported by direct numerical integrations.
\citet{deller-maddison-2005}, for instance, found a high efficiency of capturing
for a wide range of planet masses, from $0.01M_J$ to $3M_J$, assuming a solar-mass central star.
Within these limits $p_{res}$ shows rather a weak dependence on $M_p$, namely a
moderate growth with $M_p/M_\ast$ \citep{beauge-ferrazmello-1994,lazzaro-et-al-1994},
justifying our simple assumption $p_{res} = 0.5$.

Apart from a fixed $p_{res}$, we used a fixed value $e_{res}$ of 0.2 in the numerical
examples. Contrary to $p_{res}$, this quantity is not important at all for denser disks,
because the maximum eccentricity attained by the grains is limited by collisions, rather
than by resonant dynamics; see Fig.~\ref{fig_scenarioI_emax}.
That is why $e_{res}$ does not enter Eqs.~(\ref{tau dense})--(\ref{tau dense num}).
For low optical depths, $\tau$ is directly proportional to $e_{res}$,
see Eqs.~(\ref{tau tenuous})--(\ref{tau tenuous num}).
As $e_{res}$ is known to lie approximately in the range $[0.1, 0.4]$,
the uncertainty is only  a factor of two even in this case.

In the above treatment, we assumed destructive collisions to simply eliminate
both colliders. This is supported by simple estimates.
For two colliders of equal size $s$, the minimum relative
speed $v$ needed to disrupt and disperse them, is determined by
\citep[see Eq. (5.2) of ][]{krivov-et-al-2005}
\be
  v_{min} = \sqrt{8 Q_D^\ast} ,
\ee
where $Q_D^\ast$ is the minimum specific energy for fragmentation and dispersal.
At dust sizes, $Q_D^\ast \sim 10^7$ to $10^8\erg\g^{-1}$
\citep[e.g.][]{grigorieva-et-al-2006}.
This gives $v_{min} \sim 100$ to $300 \m\second^{-1}$, much smaller than
typical relative velocities (\ref{v_0}).
Under the same conditions, the largest fragment is  an order
of magnitude smaller than  either collider
\citep[see Sect. 5.3 of ][]{krivov-et-al-2005},
which allows us to ignore the generated fragments.
We also ignore cratering collisions that may only moderately affect
the mass budget, the timescales, and the optical depth of the clumps.

Perhaps the most fundamental simplification that we make is to choose
a single particle size. This leads, in particular, to
overestimation of $T_{coll}^0$ and $T_{coll}$.
The actual collisional lifetime will  be shorter,
because the grains that dominate the clumps will preferentially be destroyed
by somewhat smaller,  hence more abundant, particles.
Collisions with interstellar grains may shorten the collisional lifetime even more.

In the future, it would be interesting to develop a more elaborate kinetic model
with $n$ as a function of size $s$ 
\citep{krivov-et-al-2005,krivov-et-al-2006}.
Such a model would be  particularly interesting for dilute
disks with $\tau_0 \la 10^{-6}$, expected to be resolved with future instruments
(Herschel, ALMA, JWST, DARWIN, SAFIR, etc.)

\section{Scenario~II}

In this section, we analyze the alternative scenario~II, in which
dust parent bodies are already locked in a certain resonance, so that
the dust they produce resides in the same resonance, creating the clumps.

\begin{figure*}
  \begin{center}
  \includegraphics[scale=0.80]{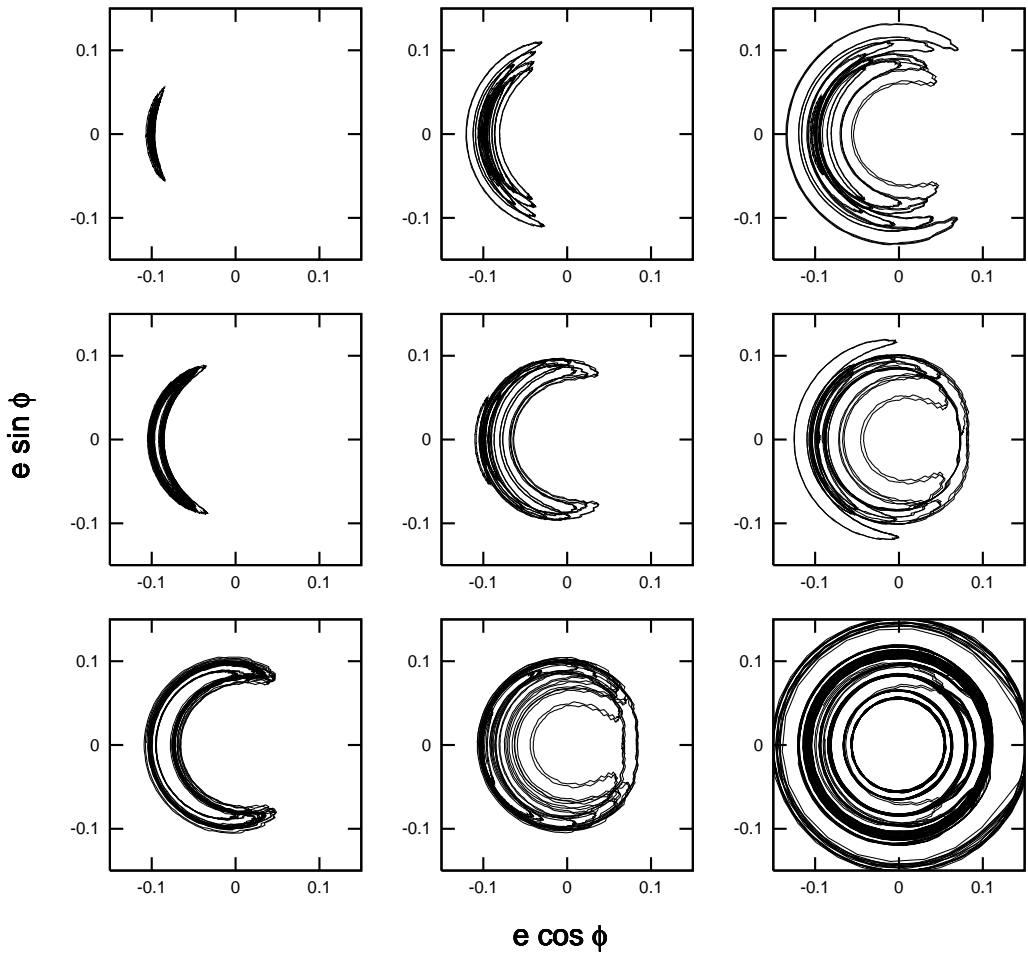} 
  \end{center}
  \caption{Phase portraits of resonant planetesimals and the dust grains they release
  in mutual collisions.
  Top: planetesimals themselves,
  middle: $40\mum$-sized dust,
  bottom: $20\mum$-sized dust.
  From left to right:
  the relative velocities of 0, 10, and $20\m\second^{-1}$. }
  \label{fig_portraits}
\end{figure*}

This scenario is more difficult to quantify than the first one, for
the following reasons.
In scenario~I, the dust production mechanisms
were ``hidden'' in one quantity, the dust injection rate $\dot{n}^+$,
which we estimated from 
an {\em ``observed''} value, optical depth $\tau_0$. This way essentially
eliminates much of the uncertainty in modeling  scenario~I.
In contrast, a physical description of scenario~II requires knowledge
of various parameters of resonant planetesimal populations. These are
not observable directly and can only be accessed by extensive modeling,
taking into account the wealth of processes that drive a forming
planetary system \citep{wyatt-2003}.
Even if the parameters of a planetesimal ensemble were known, another
major source of uncertainties would be the necessity of modeling a collisional
cascade from planetesimal-sized down to dust-sized objects, i.e. across
some 30 orders of magnitude in mass.
Therefore, we construct here the simplest possible model to explore the
efficiency of scenario~II, while being aware of its roughness.

Consider a population of planetesimals with the total mass $M$, locked in
the $(p+q):p$- resonance. These planetesimals occasionally experience
destructive collisions, in which smaller fragments are produced.
Since the relative velocities of fragments are small compared to orbital
velocities, most of them cannot leave the resonance and stay in the system.
Gradual loss of material occurs through
dust-sized fragments: below a certain minimum radius $s_{min}$, the radiation
pressure will force the particles out of the resonance
\citep{wyatt-2006}.
Relative velocities, expected to be higher for smaller fragments,
will also liberate a fraction of particles from the resonance.

We  numerically investigated the conditions under which 
the fragments remain captured in the resonance.
Again, we took the $\epsilon$~Eri system with
$M_\ast = 0.8M_\odot$, $M_p = 0.1 M_J$, $a_p = 40\AU$.
We first chose the set of planetesimal orbits, all with
the resonant (3:2) value of the semimajor axis, randomly chosen eccentricities 
between 0 and 0.1, inclinations between  0 and 0.1~rad,
and three angular elements between $0^\circ$ and $360 ^\circ$,
and selected those that turned out to be locked in resonance with
libration amplitudes not larger than $30^\circ$ (Fig.~\ref{fig_portraits} top left).
{\em From the same set of planetesimals}, we then released
dust particles with different $s$ (and $\beta$) and different velocity increments
${\bf u}$ relative to the parent bodies and numerically followed the motion of these ejecta
to find out whether they will stay in the resonance as well.
As a criterion for staying in the resonance, we simply require
that the resonant argument $\Phi$ librates with a moderate amplitude
$A$, not exceeding 30--50 degrees. We thus exclude shallow resonances that can
easily be destroyed by external perturbations and, besides, lead to nearly axisymmetric
ring-like configurations.

The results are presented in Fig.~\ref{fig_portraits}
in the form of phase portraits of eccentricity and resonant angle, in polar coordinates.
It is seen that particles larger than about $40\mum$ (or $\beta < 0.003$),
if ejected with relative velocities $u < 10\m\second^{-1}$, typically stay locked in
the resonance.
Thus $\beta_{crit} \approx 0.003$ and $u_{crit}/v_k \approx 0.003$.
The fact that $\beta_{crit}$ and $u_{crit}/v_k$ are close to each other is not surprising.
Both radiation pressure and an initial velocity ``kick'' do nothing else than cause the
initial semimajor axis and eccentricity of the particle orbit to differ from those of a parent 
planetesimal.
Within a factor of several, 
$\Delta a/a \sim e \sim \beta$ in the first case
and $\Delta a/a \sim e \sim u/v_k$ in the second.
Thus our finding means that a maximum relative change in $(a,e)$ that still leaves
the grains in the resonance is several tenths of a percent.

Interestingly, a similar problem was investigated before in quite a different
context. \citet{krivov-banaszkiewicz-2001} studied the fate of tiny icy
ejecta from the saturnian moon Hyperion, locked in the 4:3-MMR with Titan
and found that about a half of $100\mum$-sized grains remained
locked in the resonance.

One of the two effects~--- the $\beta$-threshold~--- was recently considered in great 
detail by \citet{wyatt-2006}. For the 3:2-resonance, he found
\be \label{beta crit}
  \beta_{crit} 
  = 0.034 \left( M_p/M_J \over M_\ast/M_\odot \right)^{1/2} .
\ee
For $M_\ast = 0.8M_\odot$ and $M_p = 0.1 M_J$,
this gives
$\beta_{crit} = 0.012$, which is a slightly more ``tolerant'' a threshold than the one we observed
in our simple simulation.

\begin{figure}
  \begin{center}
  \includegraphics[scale=0.34]{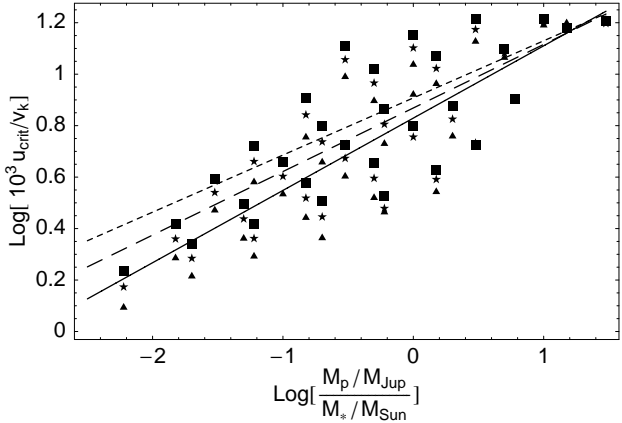} 
  \end{center}
  \caption{
  Critical relative velocity of fragments $u_{crit}/v_k$ 
  to stay in the resonance, as a function of the planet-to-star mass ratio.
  Symbols: results of individual runs; lines: power law fits (\ref{u crit}).
  Different maximum $A$ of particles were adopted as a criterion of staying in the resonance:
  $A=30^\circ$ (triangles and solid line),
  $A=40^\circ$ (squares and dashed line),
  $A=50^\circ$ (stars and dotted line).
  }
  \label{fig_ucrit}
\end{figure}

As our numerical test was done for one pair $(M_\ast , M_p)$,
it still does not allow us to judge how
the threshold of relative velocity scales with stellar mass and planetary mass.
To find an approximate scaling rule, we therefore undertook a series
of numerical integrations with
5 values of $M_p$ ($0.03$, $0.1$, $0.3$, $1.0$, and $3.0M_J$) and
6 values of $M_\ast$ ($0.1$, $0.2$, $0.5$, $1.0$, $2.0$, and $5.0M_\odot$),
$5\times6 = 30$ runs in total.
In each run, we integrated trajectories of 100 particles
$\beta = 0$ released with different velocities
from planetesimals locked in the resonance with
libration amplitudes not larger than $30^\circ$.
For each run separately, we determined the mean velocity threshold below which
the particles stay in the resonance, having libration amplitudes of not more
than $30^\circ$, $40^\circ$, and $50^\circ$ (Fig.~\ref{fig_ucrit}, symbols).
We then fitted the results with a power law
\be \label{u crit}
  u_{crit}/v_k
  = {\cal A} \left( M_p/M_J \over M_\ast/M_\odot \right)^{\cal B} .
\ee
The constants are ${\cal A} \approx 0.007 $ ($\pm 40\%$)
and ${\cal B} \approx 0.28$ ($\pm 10\%$) for the ``threshold'' $A$ of $30^\circ$.
The fits are shown in Fig.~\ref{fig_ucrit} with lines.
For the $\epsilon$~Eri under the same assumptions as above,
Eq.~(\ref{u crit}) gives $u_{crit}= 14\m\second^{-1}$
for a planet with mass $M_p = 0.1M_J$ 
and $26\m\second^{-1}$ for $M_p=1.0 M_J$.

We note that the dust grains making the
observed clumps have gone through a collisional cascade.
The cascade starts with collisions 
of the largest planetesimals with mass $m_{max}$ (or radius $s_{max}$),
followed by collisions of progressively smaller bodies,
until dust grains with mass 
$m_{min}$ (corresponds to $s_{min}$ or $\beta_{crit}$ explained above)
are produced.
We first consider a single collision and estimate
the fraction of material which has $u < u_{crit}$.
It can be approximated as \citep{nakamura-fujiwara-1991}
\be \label{Psi}
 \Psi(< u_{crit}) = 1 - \left( u_0 / u_{crit} \right)^\gamma
\qquad
 (u_{crit} \ge u_0)
\ee
where laboratory impact experiments suggest
the lower cutoff $u_0 \sim 5\m\second^{-1}$ (for dust sizes) and 
the exponent $\gamma \sim 2$ (for consolidated material, such as low-temperature ice).
Of course, the power-law fit (\ref{Psi}) should be treated with great
caution. In reality,
no sharp lower cutoff exists, so that
$u_0$ just places the lower bound on the speed range in which
Eq.~(\ref{Psi}) is valid.
A single power-law index $\gamma$ holds only in a limited velocity range.
Both $u_0$ and $\gamma$ may vary significantly for different materials,
morphology of the colliders, impact velocities, incident angles,
target and projectile masses, and other parameters of collision
\citep[e.g.][]{fujiwara-1986}.

According to Eq.~(\ref{Psi}),
most of the fragments of a single collision have velocities
slightly above $u_0$.
This is consistent with many impact experiments
that reported relative velocities in the range $5$--$20\m\second^{-1}$
as typical \citep[e.g.][]{arakawa-et-al-1995,giblin-et-al-1998}.
These typical velocities~--- and therefore
the cutoff value $u_0$~--- depend on the fragments' mass.
Indeed, \citet{nakamura-fujiwara-1991} found a mass-velocity
relation $u \propto m^{-1/6}$ to fit their experimental data.
Since all impact experiments are only possible and have only been done
for the lower end of the size distribution (objects less than one meter
in size), we will assume that $u_0 \sim 5\m\second^{-1}$ is valid for
``the last'' collisions in the cascade and that $u_0$ diminishes
according to the $m^{-1/6}$ law for ``previous'' collisions of
larger objects. Needless to say, this prescription should also be
treated with caution and is only suitable for crude estimates of the velocity-related
effects we are aiming at.

A natural question is: what is the average number of collisions, $k$,
needed to produce a dust particle with $s_{min}=30\mum$
out of parent planetesimals with size $s_{max} = 10\km$?
Simple analytical calculations done in the Appendix yield $k \approx 8$.
This result depends, albeit rather weakly, on several 
parameters, such as the mass of the largest fragment in a single binary collision.

We now estimate roughly the loss of particles due to
$u > u_{crit}$ over the whole cascade.
We simply replace the real cascade, in which each collision generates a mass distribution
of fragments, with a sequence of $k = 8$ steps.
In the first step, planetesimals of mass $m_{max}$ collide
with each other and produce smaller
fragments of equal mass. In the next step, the latter collide with each other,
and so on, until the fragments of mass $m_{min}$ are produced.
As follows from Eq.~(\ref{Psi}) and the mass-velocity relation discussed above,
the fraction of dust mass that is able to stay in the resonance
can be estimated as
\bea
 \Psi_{cascade}
 &\approx&
 \left[1 - \left( u_0              \over u_{crit} \right)^\gamma \right]
 \times
 \left[1 - \left( u_0 \delta^{1/6} \over u_{crit} \right)^\gamma \right]
\nonumber\\
 &\times&
 \ldots
 \times
 \left[1 - \left( u_0 \delta^{k/6} \over u_{crit} \right)^\gamma \right] ,
\eea
where
\be
  \delta
  \equiv 
  \left( m_{min} \over m_{max} \right)^{1/k}
  \ll 1 .
\ee
In nearly all situations,
only the first term matters, which means
that most of the material is lost to the ``velocity effect'' at the
last collision. In what follows, we thus adopt
\bea
 \Psi_{cascade}
 &\approx&
 1 - \left( u_0              \over u_{crit} \right)^\gamma .
\eea

On the whole, the resonant population
represents a nearly closed system, losing material only slowly through the smallest
and fastest grains, which reduces the mass of the population on extremely long 
(typically Gyr) timescales. This makes it possible to apply
 Dohnanyi's (\citeyear{dohnanyi-1969})
theory of a closed collisional system. Provided that the critical impact energy
for fragmentation and dispersal $Q_D^\ast$ is independent of size over the size regime
considered, he has shown that such a system evolves to a steady state,
in which the mass distribution is a simple power law:
$n(m) \propto m^{-1-\alpha}$ with $\alpha = 5/6 = 0.83$.
\citet{durda-dermott-1997} generalized this result by considering a more
realistic, size-dependent $Q_D^\ast$.
They have shown that a power-law dependence of
$Q_D^\ast$ on size preserves a power-law size distribution, but makes
it steeper or flatter, depending on whether $Q_D^\ast$ decreases or increases with size.
For $Q_D^\ast \propto s^{-0.3}$,
which is a plausible choice in the strength regime
\citep[see][and references therein]{krivov-et-al-2005},
Durda and Dermott found $\alpha = 0.87$.

The total cross section of the steady-state population is
\be \label{Sigma}
 \Sigma =
 {1-\alpha \over \alpha - 2/3}
 \pi^{1/3} \left( 3 \over 4\rho \right)^{2/3}
 \!\!\!\!\!\!
 {
  M 
   \over
  m_{max}^{1-\alpha}\;\; m_{min}^{\alpha - 2/3}
 }
 \Psi_{cascade} ,
\ee
where $\rho$ is the material density of the objects.
For $\alpha =5/6$, Eq.~(\ref{Sigma}) takes a symmetric form
\be \label{Sigma Dohnanyi}
 \Sigma =
 \pi^{1/3} \left( 3 \over 4\rho \right)^{2/3}
 \!\!\!\!\!\!
 {M \over m_{min}^{1/6} \;\; m_{max}^{1/6}}
 \Psi_{cascade} .
\ee
Equations (\ref{Sigma}) and (\ref{Sigma Dohnanyi}) show that
the results depend on $m_{min}$ and $m_{max}$ only weakly.

\begin{figure}
  \begin{center}
  \includegraphics[scale=0.32]{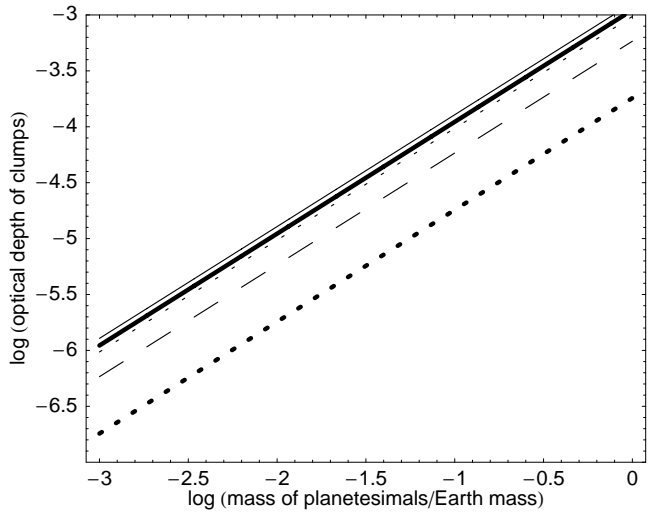} 
  \end{center}
  \caption{Normal optical depth of the resonant clumps $\tau$
  in scenario~II,
  as a function of the planetesimal population mass $M$.
  Thick lines are for $u_0=5\m\second^{-1}$ and $\gamma=2$ and different $\alpha$:
  Dohnanyi's $\alpha = 0.833$ (dashed) and
  a more realistic $\alpha = 0.87$ (solid).
  Thin lines are for $\alpha = 0.87$ and different $u_0$ and $\gamma$:
  $u_0=1\m\second^{-1}$ and $\gamma=2$ (solid),
  $u_0=10\m\second^{-1}$ and $\gamma=2$ (dashed),
  $u_0=5\m\second^{-1}$ and $\gamma=1.5$ (dotted).
  Cf. Fig.~\ref{fig_scenarioI} for scenario~I.
  }
  \label{fig_scenarioII}
\end{figure}

Like in the first scenario, we then divide $\Sigma$ by the
area of the clumps, $S = 4 \pi \hat{S} a^2 e_{max}$,
to obtain an estimate of the clumps' normal optical depth.
The results are presented in Fig.~\ref{fig_scenarioII}
for the $\epsilon$~Eri case.
We assumed again $M_\ast = 0.8 M_\odot$, $\epsilon=0.1=6^\circ$, and
considered the 3:2 resonance (planetesimals with $a=56$~AU)
with a planet of mass $M_p = 0.1M_J$.
We adopted $s_{max}=10\km$, $e_{max} = 0.2$, and $\hat{S} = 0.2$.
From (\ref{beta crit}),
$\beta_{crit} = 0.012$ (or $s_{min} = 12\mum$).
As in Fig.~\ref{fig_scenarioI}, we plot the normal optical depth
of the clumps $\tau$, but now as a function of the planetesimal population mass $M$.
Two size-distribution indexes were used: Dohnanyi's $\alpha = 0.833$ and
a more realistic $\alpha = 0.87$.
Our best-guess choice, $\alpha = 0.87$  shows that the population
with $M = 0.01M_\oplus$
($M_\oplus$ is the Earth mass)
would produce clumps with $\tau \sim 10^{-5}$.
The same figure demonstrates that the dependence on $\gamma$
is moderate, but that on the cutoff value $u_0$ is rather substantial.
We speculate that uncertainties in the fragment velocity distribution
may introduce at least an order of magnitude, perhaps a larger, uncertainty
in the resulting optical depth of the clumps.

\begin{figure*}[htb!]
  \begin{center}
  \includegraphics[scale=0.33]{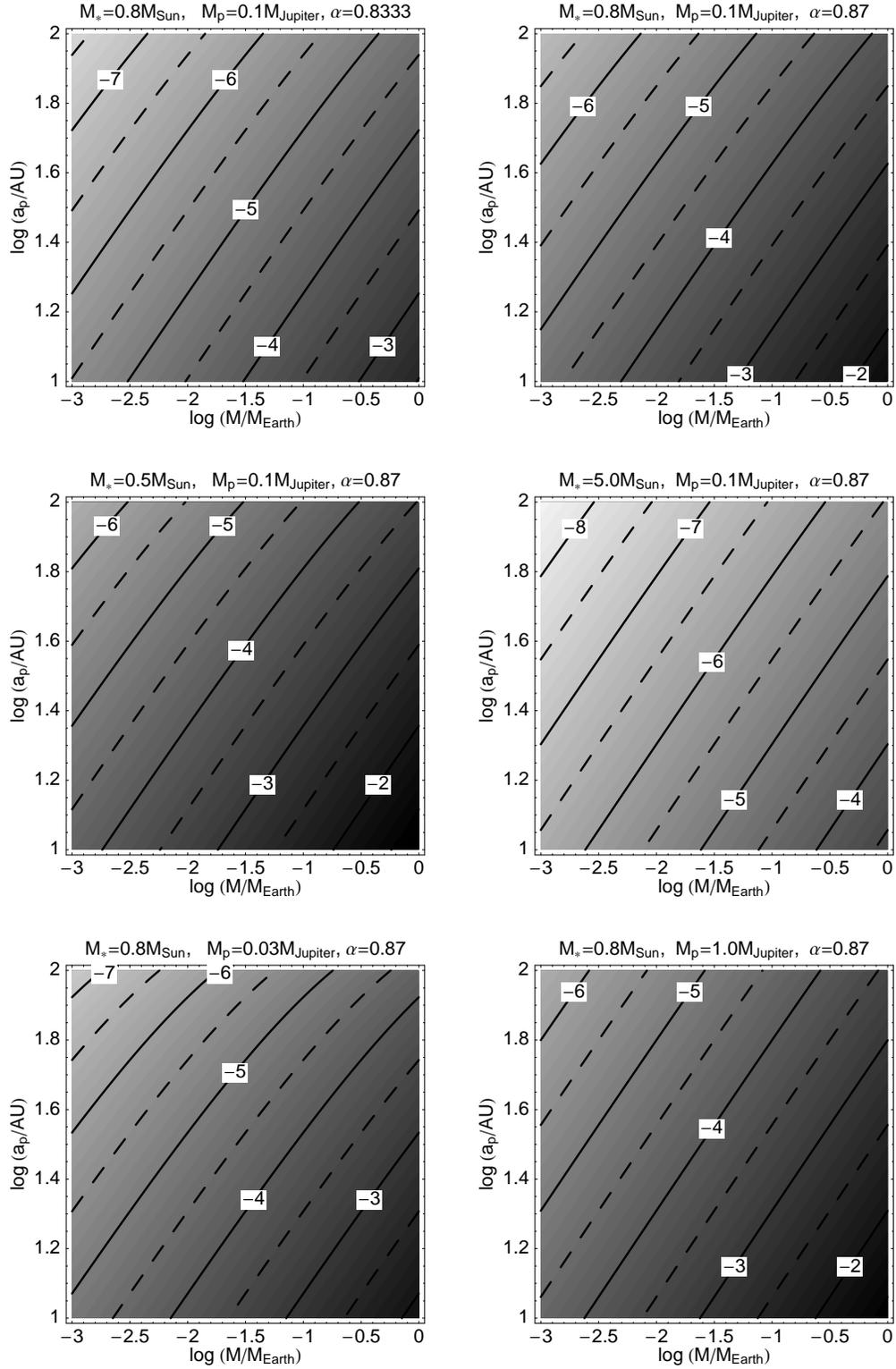} 
  \end{center}
  \caption{Contour plots of the normal optical depth of the resonant clumps $\tau$
  in scenario~II,  as a function of the planetesimal population mass $M$ and
  the planet's distance from the star.
  Top left and top right:    Dohnanyi's $\alpha = 0.833$ and
  more realistic $\alpha = 0.87$, both for the same star with $0.8M_\odot$
  and a planet with 0.1 Jupiter mass.
  Middle left and middle right: stars with 0.5 and $5.0M_\odot$, both
  for $\alpha = 0.87$  and a planet with 0.1 Jupiter mass.
  Bottom left and bottom right:
  a planet with 0.03 Jupiter mass and 1 Jupiter mass, both for
  $\alpha = 0.87$ and $M_\ast = 0.8M_\odot$.
  Contours are labeled with the values of $\log\tau$.
  Cf. Fig.~\ref{fig_scenarioI_plots} for scenario~I.}
  \label{fig_scenarioII_plots}
\end{figure*}

Similar to the first scenario, we now investigate the dependence
of the clump formation efficiency on various parameters.
We use Eq.~(\ref{Sigma}) with $\alpha=0.87$ and $\rho=2\g\cm^{-3}$
and express $m_{min}$ through $\beta_{crit}$ and then through
$M_\ast$ and $M_p$ by Eq.~(\ref{beta crit}).
By setting  $\Psi_{cascade}$ to unity, we obtain a rough upper
limit on the optical depth of the clumps:
\bea \label{tau scII num}
  \tau &\la&
     4.3\cdot 10^{-5}
     \left(\frac{\hat{S}}{0.2}\right)^{-1}
     \left(\frac{e_{max}}{0.2}\right)^{-1}
\nonumber\\
&\times&
     \left(\frac{s_{max}}{10\;\mathrm{km}}\right)^{-0.39}
     \left(\frac{M}{0.1 M_\oplus} \right)
\nonumber\\
&\times&
     \left(\frac{M_p}{M_J}\right)^{0.31}
     \left(\frac{M_\ast}{M_\odot}\right)^{-2.14}
     \left( \frac{a_p}{100\AU} \right)^{-2}
     \left(\frac{p}{p + q}\right)^{4/3} .
\eea

Each panel in Fig. \ref{fig_scenarioII_plots} presents the results in form of the contour plots
of equal $\tau$, the average optical depth of the clumps, on the total mass of
the resonant planetesimals $M$ and planet's orbital radius $a_p$,
showing that $\tau$ grows with increasing $M$ and decreasing $a_p$.
Different panels illustrate the dependence on $\alpha$, $M_\ast$, and $M_p$.
As  already seen, the dependence on $\alpha$ is quite strong.
The dependence on the stellar mass is strong, too. However, in contrast to
the first scenario, stars with higher luminosity produce less pronounced clumps. 
Dependence on the planet mass $M_p$ is stronger than in the first scenario.
More massive planets create brighter clumps.

Remember that all our estimates imply that the resonant population is in a steady 
state.
Transient fluctuations in the population can be produced by recent major collisional
events \citep{wyatt-dent-2002,kenyon-bromley-2005,grigorieva-et-al-2006}.
Usually, the dust debris cloud generated by a ``supercollision'' of two large
planetesimals is spread by the Keplerian shear 
into a relatively homogeneous ring 
in only $10^4$ to $10^5$ years \citep{kenyon-bromley-2005}.
However, in a resonant population, most of the fragments should stay in the resonance and 
contribute to the clump brightness~--- for the reasons discussed above.
The same 
Eqs.~(\ref{Sigma}) and (\ref{tau scII num}) apply,
with $M$  now the total mass 
of both colliders.
Roughly, a collision of two lunar-sized objects would generate $0.01M_\oplus$ of debris,
enough to produce well-observable clumps.

\section{Conclusions and discussion}

In this paper, we address the possible origin of azimuthal substructure
in the form of clumps or blobs, observed in some debris disks such as $\epsilon$~Eridani.
Two scenarios are considered. In the first case, the one discussed most in the literature,
dust particles are produced in planetesimal belts exterior to resonance locations,
brought to exterior planetary resonances by the P-R effect, and their resonance
trapping leads to the formation of clumps.
In the second case, dust-producing planetesimals
are already locked in a resonance, for instance, as a result of planetary migration
in the past, and the dust they produce resides in the same resonance, creating the clumps.
For both scenarios, we have constructed simple analytic models in order to explore
their applicability limits and efficiency for a wide range of stars,
planets, disks, and planetesimal families.

{\em 1. Scenario~I.}
We conclude that the efficiency of the first scenario strongly depends on
the normal geometrical optical depth $\tau_0$ of the ``regular'' disk, from which
dust is supplied to the clumps. 
The optical depth is known from observations.
Besides, the brightness of the resulting clumps is nearly proportional to
the probability of trapping ``visible'' dust grains in the resonance, $p_{res}$.
This parameter can be constrained relatively well, too,
with the aid of numerical integrations
for one or another particular system and a specified resonance.

In disks with roughly $\tau_0 \la 10^{-4}$--$10^{-5}$, the first scenario
works well, creating pronounced clumps.
Their optical depth $\tau$ is usually lower than, but can be comparable to or even exceed,
the optical depth of the background disk. In other words, in dilute disks the clumps can be 
almost as bright as, and sometimes brighter than, the underlying disk in the same region.
This is the case in the solar system's interplanetary dust cloud
and the dust in the Kuiper belt with their $\tau_0 \la 10^{-7}$. 
Further, this will be the case for the debris disks at which future telescopes
are aiming;
examples are:
the Herschel Space Observatory,
the Atacama Large Millimeter Array (ALMA),
the James Webb Space Telescope (JWST),
the space infrared interferometer DARWIN,
and the Spitzer's and Herschel's follow-on SAFIR.
But it is not the case for debris disks of the other stars resolved so far.
The disks currently observed and the clumps observed in them,
all have optical depth $> 10^{-5}$.

For higher optical depths $\tau_0$, $\tau$  can still be high enough,
but collisions are ``killing'' the clumps in the following way.
They shorten the lifetimes of grains captured in the resonance, which prevents them
from developing higher orbital eccentricities. Instead, the trapped particles
remain in orbits with low eccentricities. The resonant population will look like a narrow 
ring without azimuthal structure, not like clumps. The dustier the disk, the
narrower the resulting ring (for planets in near-circular orbits),
and the less pronounced the azimuthal structure.
For dustier disks, we find that the optical depth of the clumps has a tendency 
to increase when the responsible planet orbits closer to the star and has a higher mass.
It also tends to grow with the luminosity of the central star.
Increasing the opening angle of the dust disk has the same effect.
However, all these dependencies are quite weak.

In the case of $\epsilon$ Eridani, which falls into the category of ``dustier'' disks,
the first scenario does not seem to be consistent
with the presence of several longitudinally confined ``blobs''.

{\em 2. Scenario~II.}
The efficiency of the second mechanism is obviously proportional to the total mass
of the resonant population of planetesimals, but is strongly affected by other
parameters, too.
The brightness of the clumps produced by the second scenario increases with 
the decreasing luminosity of the star, increasing planetary mass, and decreasing orbital radius
of the planet. All these dependencies are more pronounced than in the first scenario.

In the case of $\epsilon$ Eridani, a family with a total mass of $\sim 0.01$ to $0.1$
Earth masses could well be sufficient
to account for the observed optical depth of
the clumps, roughly $\tau \sim 10^{-5}$. Thus this mechanism
appears to be more relevant for systems like $\epsilon$~Eri.

It must be noted that the efficiency of scenario~II
is quite sensitive to the poorly known properties of the collisional grinding process.
One is the critical energy for fragmentation and dispersal $Q_D^\ast$ over the whole
range of masses~--- from planetesimals down to dust~--- which strongly affects
the equilibrium slope of the size distribution, and therefore the total cross section
ares of the resonant clumps.
Even more important is the distribution of the ejecta velocities in disruptive collisions,
which determines what fraction of collisional debris is fast enough to leave the resonant 
population.
For these reasons, models of the second scenario are quantitatively more uncertain
than those of the first one.

Models presented here are the simplest possible ones we could construct.
Accordingly, they involve a wealth of simplifying assumptions and therefore
cannot substitute for developing more elaborate models.
Yet we hope that our results can give useful guidelines
and serve as a starting point for more detailed
studies of clumps in debris disks.

\acknowledgements
{
We wish to thank Mark Wyatt and Jane Greaves for useful discussions and the anonymous referee
for his/her thorough and very helpful review.
We are grateful to Martin Reidemeister 
for his assistance with numerical runs of the integration
routine.
Miodrag Srem{\v c}evi{\' c} is supported by the Cassini UVIS project.
}

\renewcommand{\d}{\ensuremath{\mathrm{d}}}

\appendix
\section{Number of collisions over the collisional cascade}

Consider the collisional cascade that produces small grains of mass $m$ from parent bodies 
of mass $m_{max}$. 
We wish to estimate the mean number of collisional events that the grains of mass $m$
must have had in their collisional history.

Each individual fragmenting collision of this cascade can be described by a distribution 
$g(m, m_x)$ of fragments $m$ from the largest one $m_x$ down to infinitely small ones.
Let the normalization condition be given by
$$ \int_0^{m_x}\limits g(m, m_x) \mathrm{d}m = 1.$$
Now we introduce a relation between $m_x$ and $m_{max}$ given by $\mu = m_x / m_{max}$.
As each catastrophic collision in a cascade reduces the mass of the largest possible fragment 
by the factor $\mu$, the maximum number of collisions that a fragment of size $m$ might have 
undergone since the parent body $m_{max}$ was disrupted is given by
$$
  k_\mathrm{max} = \frac{ \ln(m/m_{max}) }{\ln\mu}.
$$
If we assume every collision to be described by the same $\mu$ and the same power-law dependence for
$$ g(m, m_x) \equiv (2 - \eta) \frac{m_x^{\eta - 2}}{m^{\eta - 1}}, $$
but with a different absolute value of $m_x$,
the resulting distribution after $k+1$ collisions is given by
\begin{eqnarray}
  P_k(m, m_{max})
  &=&
  \int_{m\mu^{-k}}^{m_{max}\mu}\limits   \!\!\!\!\! g(m_1, m_{max}\mu)
  \int_{m\mu^{1-k}}^{m_1\mu}\limits      \!\!\!\!\! g(m_2, m_1\mu)
  \int_{m\mu^{2-k}}^{m_2\mu}\limits      \!\!\!\!\! g(m_3, m_2\mu)
  \cdots\nonumber\\
  &&
  \int_{m\mu^{-2}}^{m_{k - 1}\mu}\limits g(m_k, m_{k - 1}\mu) \; g(m, m_k\mu)\:
  \nonumber\\
  &&
  \d{m_k}\cdots\d{m_3}\d{m_2}\d{m_1},
\end{eqnarray}
where $P_k$ provides the fraction of mass that falls in the range $[m,m+\d{m}]$ after $k + 1$ 
collisions.
If we change to logarithmic scaling of the integration variables, i.e.
$\d{m_1} = m_1 \d{(\ln\frac{m_1}{m})}$ and so on, and explicitly insert $g$,
we obtain 
\begin{eqnarray}
  P_k(m, m_{max}) 
  &=& (2 - \eta)^{k + 1} 
  \int_{\ln \mu^{-k}}^{\ln\frac{\mu m_{max}}{m}}\limits \frac{(m_{max} \mu)^{\eta - 2}}{m_1^{\eta - 1}}\:
  \int_{\ln \mu^{1-k}}^{\ln\frac{\mu m_1}{m}}\limits \frac{(m_1 \mu)^{\eta - 2}}{m_2^{\eta - 1}}\nonumber\\
  &&
  \int_{\ln \mu^{2-k}}^{\ln\frac{\mu m_2}{m}}\limits \frac{(m_2 \mu)^{\eta - 2}}{m_3^{\eta - 1}}\:
  \cdots
  \int_{\ln \mu^{-2}}^{\ln\frac{\mu m_{k - 1}}{m}}\limits \frac{(m_{k - 1}\mu)^{\eta - 2}}{m_k^{\eta - 1}} \;
     \frac{(m_k \mu)^{\eta - 2}}{m^{\eta - 1}}\nonumber\\
  &\times& m_k\cdots m_3 m_2 m_1 \nonumber\\
  &&\d{(\ln \frac{m_k}{m})}\cdots\d{(\ln \frac{m_3}{m})}\d{(\ln \frac{m_2}{m})}\d{(\ln \frac{m_1}{m})}.
\end{eqnarray}
Luckily, all integration variables disappear and the integrand is constant:
\begin{eqnarray}
  P_k(m, m_{max}) &=& (2 - \eta)^{k + 1} \frac{1}{m}\left(\frac{\mu^{k + 1} m_{max}}{m} \right)^{\eta - 2} 
  \nonumber\\
  &\times&
  \int_{\ln \mu^{-k}}^{\ln\frac{\mu m_{max}}{m}}\limits\:
  \int_{\ln \mu^{1-k}}^{\ln\frac{\mu m_1}{m}}\limits\:
  \int_{\ln \mu^{2-k}}^{\ln\frac{\mu m_2}{m}}\limits\:
  \cdots \int_{\ln \mu^{-2}}^{\ln\frac{\mu m_{k - 1}}{m}}\limits\nonumber\\
  && \d{(\ln \frac{m_k}{m})}\cdots\d{(\ln \frac{m_3}{m})}\d{(\ln \frac{m_2}{m})}\d{(\ln \frac{m_1}{m})}.
\end{eqnarray}
Now the variables are shifted starting with $m_1 \rightarrow \mu^k m_1$:
\begin{eqnarray}
  P_k(m, m_{max}) &=& (2 - \eta)^{k + 1} \frac{1}{m}\left(\frac{\mu^{k + 1} m_{max}}{m} \right)^{\eta - 2}
  \nonumber\\
  &\times&\int_0^{\ln\frac{\mu^{k + 1} m_{max}}{m}}\limits\:
  \int_{\ln \mu^{1-k}}^{\ln\frac{\mu^{1-k} m_1}{m}}\limits\:
  \int_{\ln \mu^{2-k}}^{\ln\frac{\mu m_2}{m}}\limits\:
  \cdots \int_{\ln \mu^{-2}}^{\ln\frac{\mu m_{k - 1}}{m}}\limits\nonumber\\
  && \d{(\ln \frac{m_k}{m})}\cdots\d{(\ln \frac{m_3}{m})}\d{(\ln \frac{m_2}{m})}\d{(\ln \frac{m_1}{m})}.
\end{eqnarray}
Substituting everything from $m_1$ to $m_k$ in the same manner, we obtain
\begin{eqnarray}
  P_k(m, m_{max}) &=& (2 - \eta)^{k + 1} \frac{1}{m}\left(\frac{\mu^{k + 1} m_{max}}{m} \right)^{\eta - 2}
  \nonumber\\
  &\times&\int_0^{\ln\frac{\mu^{k + 1} m_{max}}{m}}\limits\:
  \int_0^{\ln\frac{m_1}{m}}\limits\:
  \int_0^{\ln\frac{m_2}{m}}\limits\:
  \cdots \int_0^{\ln\frac{m_k}{m}}\limits\nonumber\\
  && \d{(\ln \frac{m_k}{m})}\cdots\d{(\ln \frac{m_3}{m})}\d{(\ln \frac{m_2}{m})}\d{(\ln \frac{m_1}{m})}.
\end{eqnarray}
The integral equates the volume spanned by
$0 < x_k < \cdots < x_3 < x_2 < x_1 < x_0 = \ln\frac{\mu^{k + 1} m_{max}}{m}$, which is given by
\begin{eqnarray}
  && \int_0^{x_0=\ln\frac{\mu^{k + 1} m_{max}}{m}}\limits\:
  \int_0^{x_1}\limits\:
  \int_0^{x_2}\limits\:
  \cdots \int_0^{x_k}\limits
  \: \d{x_k}\cdots\d{x_3}\d{x_2}\d{x_1}\nonumber\\
  &=& \frac{\left[\ln\frac{\mu^{k + 1} m_{max}}{m}\right]^k}{k!}.
\end{eqnarray}

\begin{figure}[hbt!]
  \centering
  \includegraphics[width=\linewidth]{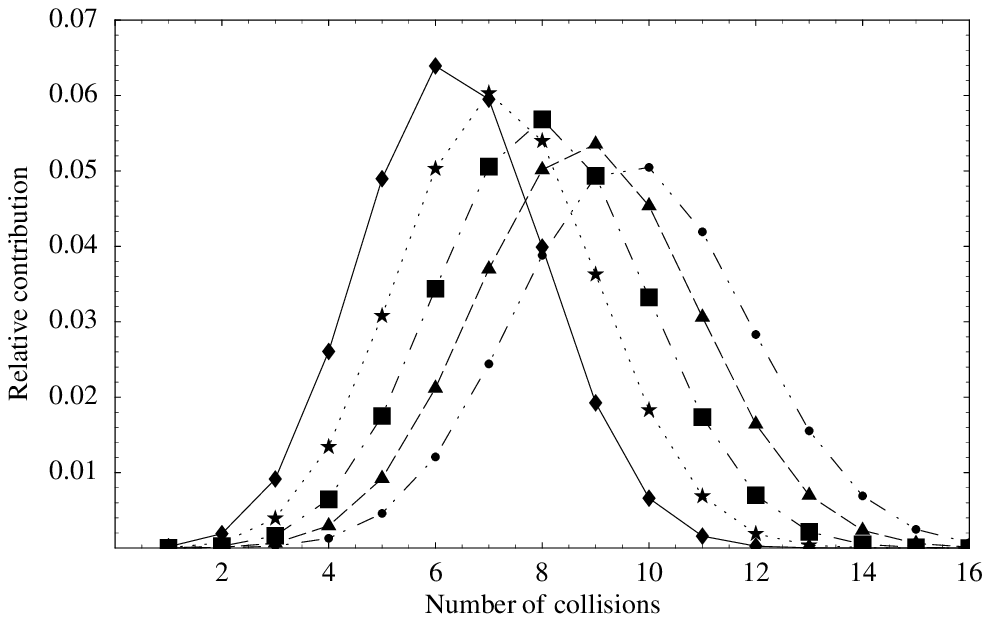}
  \includegraphics[width=\linewidth]{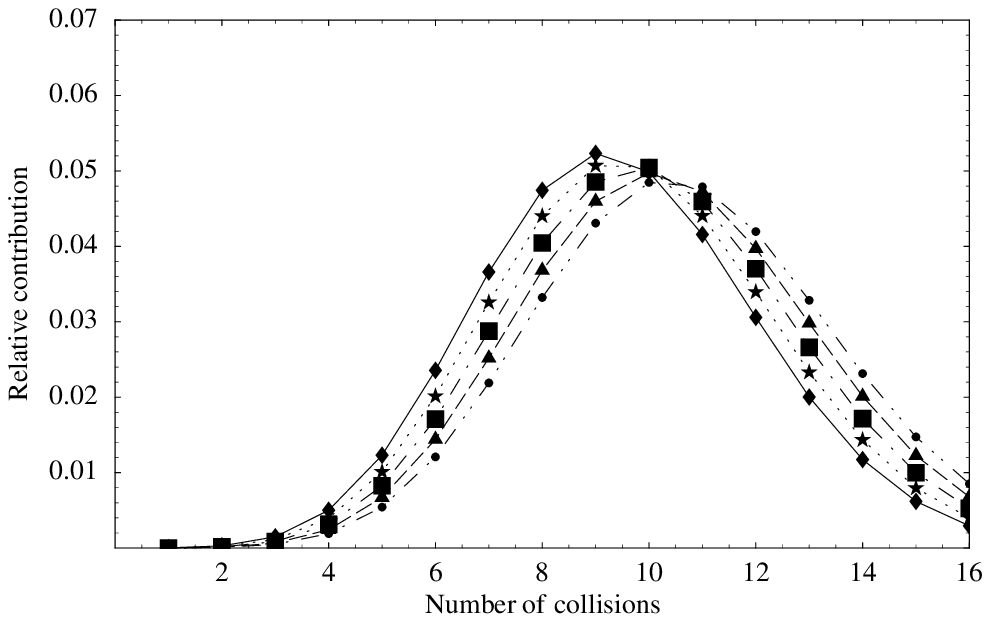}
  \caption{Probability distribution of particles with size $m$ originating from parent bodies of mass
  $m_{max} = 10^{27 + x} m$ over the number of collisions they underwent.
  $x$ has values $-6$, $-3$, $0$, $3$, $6$ (peaks 
  from left to right).
  $\eta = 11/6$. The peak position is roughly proportional to 
  $\log_{10}(m_{max}/m)/3.5$. (The lines just connect adjacent points for illustrative purpose, although the 
  distribution is discrete in $k$. The distribution looks similar to a Poisson distribution but is 
  different.)
  Top: $m_x/m_{max} = \mu = 0.1$,
  bottom: $\mu = 0.5$.}
  \label{figColCasc1}
\end{figure}

The final result is
\begin{eqnarray}
  && P_k(m, m_{max})
  \nonumber\\
  && = \frac{1}{m} (2 - \eta)^{k + 1} \left(\frac{\mu^{k + 1} m_{max}}{m} \right)^{\eta - 2} 
  \frac{\left[\ln\left(\frac{\mu^{k + 1} m_{max}}{m}\right)\right]^k}{k!}
\end{eqnarray}
or
\begin{eqnarray}
  && P_k(m, m_{max})
  \nonumber\\
  && = \frac{2 - \eta}{m} \left(\frac{\mu^{k + 1} m_{max}}{m} \right)^{\eta - 2}
  \frac{\left\{\ln\left[\left(\frac{\mu^{k + 1} m_{max}}{m}\right)^{2 - \eta}\right]\right\}^k}{k!}.
\end{eqnarray}
For $k = 0$, i.e. no intermediate collision, we are back at
\begin{eqnarray}
  P_0(m, m_{max}) &=& g(m, \mu m_{max}).
\end{eqnarray}
To obtain the distribution of the total mass of fragments per mass decade, we have to evaluate
\begin{eqnarray}
  \tilde{P} &=& m\cdot \ln (10) \cdot P.
\end{eqnarray}
Figure \ref{figColCasc1} shows examples for the distribution.
The planetesimal radius $s_{max} = 10\km$ and the grain radius $s=100$ to $10\mum$
correspond to the mass ratio  $\log_{10}(m_{max}/m) \sim 24$ to $27$,
or $x$ from $-3$ to $0$. In this case, the distributions peak at $k = 7$ to 10.
Therefore, we adopt $k = 8$ for the  bulk of the calculations.

\newcommand{\AAp}      {Astron. Astrophys.}
\newcommand{\AApSS}    {AApSS}
\newcommand{\AApT}     {Astron. Astrophys. Trans.}
\newcommand{\AdvSR}    {Adv. Space Res.}
\newcommand{\AJ}       {Astron. J.}
\newcommand{\AN}       {Astron. Nachr.}
\newcommand{\ApJ}      {Astrophys. J.}
\newcommand{\ApSS}     {Astrophys. Space Sci.}
\newcommand{\ARAA}     {Ann. Rev. Astron. Astrophys.}
\newcommand{\ARevEPS}  {Ann. Rev. Earth Planet. Sci.}  
\newcommand{\BAAS}     {BAAS}
\newcommand{\CelMech}  {Celest. Mech. Dynam. Astron.}
\newcommand{\EMP}      {Earth, Moon and Planets}
\newcommand{\EPS}      {Earth, Planets and Space}
\newcommand{\GRL}      {Geophys. Res. Lett.}
\newcommand{\JGR}      {J. Geophys. Res.}
\newcommand{\MNRAS}    {Mon. Not. Roy. Astron. Soc.}
\newcommand{\PSS}      {Planet. Space Sci.}
\newcommand{\SSR}      {Space Sci. Rev.}

\end{document}